\documentclass[sigconf,screen]{acmart}

\setcopyright{acmlicensed}
\copyrightyear{2026}
\acmYear{2026}
\acmDOI{XXXXXXX.XXXXXXX}
\acmConference[CCS '26]{ACM Conference on Computer and Communications Security}{November 15-19, 2026}{The Hague, The Netherlands}
\acmBooktitle{CCS '26: 2026 ACM SIGSAC Conference on Computer \& Communications Security Nov. 15--19, 2026, The Hague, The Netherlands}
\acmISBN{978-1-4503-XXXX-X/2018/06}

\usepackage[english]{babel}
\usepackage{balance}
\usepackage{graphicx}
\usepackage{tikz}
\usepackage{multirow}
\usepackage{multicol}
\usepackage{booktabs}
\usepackage{url}
\usepackage{arydshln}
\usepackage{enumitem}
\usepackage{caption}
\usepackage{hyperref}
\usepackage{listings}
\usepackage{mathtools}
\usepackage{amsmath}
\usepackage{xcolor}
\usepackage[normalem]{ulem}
\usepackage[english]{babel}
\usepackage[flushleft]{threeparttable}
\usepackage[all]{nowidow}
\usepackage{soul}
\usepackage{ListStyle}
\usepackage{bm}
\usepackage{tablefootnote}
\usepackage{threeparttable}
\usepackage{array, makecell}
\usepackage{colortbl}
\usepackage{amsfonts}
\usepackage{float}
\usepackage{wrapfig}
\usepackage{pifont}  
\usepackage{natbib}

\renewcommand{\thetable}{\arabic{table}}

\definecolor{newGreen}{rgb}{0.0,0.63,.333}
\definecolor{newYellow}{rgb}{1.0,.7,.1}

\useunder{\uline}{\ulined}{}%
\DeclareUrlCommand{\bulurl}{}

\newcommand{\mytool}{{ICFlowNet}}

\def\halfcheck{\textcolor{newYellow}{\checkmark\kern-1.1ex\raisebox{.7ex}{\rotatebox[origin=c]{125}{--}}}}

\colorlet{tableheadcolor}{gray!25} 
\colorlet{tablerowcolor}{gray!10} 
\begin{document}

\title[ICFlowNet]{Long-Range Indirect Control-Flow Prediction in Stripped Binaries via Dual Virtual Hubs and Multi-Task Graph Learning}


\author{Kun Liu}
\affiliation{%
  \institution{Tulane University}
  \city{New Orleans}
  \state{Louisiana}
  \country{USA}
}
\email{kliu14@tulane.edu}

\author{Zhengming Ding}
\affiliation{%
  \institution{Tulane University}
  \city{New Orleans}
  \state{Louisiana}
  \country{USA}
}
\email{zding1@tulane.edu}

\author{Chenke Luo}
\affiliation{%
  \institution{Tulane University}
  \city{New Orleans}
  \state{Louisiana}
  \country{USA}
}
\email{cluo6@tulane.edu}

\author{Tianyi Xu}
\affiliation{%
  \institution{Tulane University}
  \city{New Orleans}
  \state{Louisiana}
  \country{USA}
}
\email{txu9@tulane.edu}

\author{Zizhan Zheng}
\affiliation{%
  \institution{Tulane University}
  \city{New Orleans}
  \state{Louisiana}
  \country{USA}
}
\email{zzheng3@tulane.edu}

\author{Haotian Zhang}
\affiliation{%
  \institution{New Jersey Institute of Technology}
  \city{Newark}
  \state{New Jersey}
  \country{USA}
}
\email{haotian.zhang@njit.edu}

\author{Jiang Ming}
\affiliation{%
  \institution{Tulane University}
  \city{New Orleans}
  \state{Louisiana}
  \country{USA}
}
\email{jming@tulane.edu}

\begin{abstract}
Recovering indirect control-flow (ICF) edges is fundamental to binary security analysis, yet existing methods struggle with long-range dependencies, isolate different ICF types, and are often evaluated under protocols vulnerable to label noise and data leakage. We present \textit{ICFlowNet}, a unified framework for long-range ICF prediction in stripped binaries. ICFlowNet introduces candidate-aware Dual Virtual Hubs---a Global Code Hub and a Global Data Hub---to create short routing paths between distant code and data evidence, and combines them with multi-task graph learning to jointly model indirect calls, indirect tail calls, jump tables, and returns. To enable credible evaluation, we further develop a leakage-aware, noise-controlled pipeline with package-level splits, function-level mnemonic-hash deduplication, and a clean test protocol built from dynamic positives and absolute negatives. Using this pipeline, we construct a dataset of $15,901$ unique stripped x86\_64 binaries, including $1,351$ with dynamic ground truth. Experiments show that simply scaling static supervision yields only marginal gains, whereas our structural and multi-task designs are essential: Dual Virtual Hubs improve long-range F1 by up to 9.13 points, multi-task learning adds up to 5.81 points, and the final model outperforms prior baselines by more than 13 F1 on long-range indirect calls while adding only 11.44\% topological overhead.	
\end{abstract}

\maketitle

\section{Introduction}
\label{sec:Introduction}

Executable code forms the invisible backbone of modern computing, powering everything from personal devices and IoT systems to large-scale cloud services and sophisticated malware. Yet, in many security settings, analysts must reason about binaries without source code. A central challenge in this setting is recovering the control-flow graph (CFG), which serves as the structural foundation for many downstream analyses and defenses. Among all CFG edges, indirect control-flow (ICF) transfers---whose targets are computed dynamically at runtime---remain particularly difficult to resolve statically. This uncertainty directly weakens downstream applications such as binary rewriting~\cite{Gregory20,Sushant20,Meng21}, recompilation~\cite{Hasabnis16,BinRec20,Egalito}, software hardening~\cite{Kuznetzov14,miTrimmer,Priyadarshan23,PXoM}, and binary diffing~\cite{David17,David18,BinTuner,DeepBinDiff}.

ICF transfers arise from diverse programming constructs and compiler optimizations~\cite{Andriesse16,Meng16,Burow2017CFI,pang2020sok}. Representative cases include indirect calls through function pointers, indirect tail calls that collapse call-return pairs into jumps, jump tables used for multi-way dispatch, and \texttt{ret} instructions whose destinations depend on dynamic call history. When source code is available, compilers retain rich semantic cues such as types, function boundaries, and structural layouts that make these transfers comparatively easier to recover. In stripped binaries, however, those cues are largely absent, forcing analysis tools to infer dynamic targets from low-level instruction patterns and incomplete structural evidence.

Existing approaches remain fragmented. Traditional binary analysis tools~\cite{hexrays_idapro,nsa_ghidra,Yan16} and classical control-flow integrity techniques~\cite{zhang13,van2015practical,TypeArmor,BPA} recover only a limited subset of dynamic targets, a limitation also highlighted by Callee~\cite{Callee}. 
Recent learning-based methods improve individual subproblems, but they still operate in isolation: Callee~\cite{Callee} and AttnCall~\cite{AttnCall2024} focus exclusively on indirect-call prediction and rely on sliced, linearized context representations rather than explicit whole-program graph structure, which can limit their ability to preserve fine-grained structural relations such as code-data cross-references. Their evaluation protocols also do not enforce the stricter package-level isolation and deduplication scheme that we adopt later to better control in-package bias~\cite{allamanis2019}.
SJA~\cite{nguyen2024}, a strong jump-table analyzer, is evaluated only on an outdated LLVM 6.0 toolchain whose emitted patterns are substantially simpler than those of modern binaries. By contrast, CupidCall~\cite{NeuCall} shows that graph-based representations enriched with code-data cross-references can better preserve structural dependencies and outperform sequence-based modeling on indirect-call prediction. These advances suggest that graph learning is a promising foundation for binary ICF prediction, but a practical solution must still overcome four key challenges.

\vspace{2pt}
\noindent
\textbf{\textit{Challenge 1.}} 
Even graph-based models degrade when the evidence governing an ICF transfer lies many hops away from where the transfer is executed, because standard message-passing GNNs have a limited effective receptive field and are known to struggle with long-range dependencies due to graph bottlenecks and over-squashing~\cite{alon2021on,Topping2022Oversquashing,dwivedi2022LRGB}.
Although CupidCall~\cite{NeuCall} uses code-data cross-reference (xref) edges to expose data-flow evidence, static analysis cannot reliably guarantee that such bridges are recovered close to the relevant endpoints. In practice, critical pointer initializations may be missing entirely or may only appear several hops away from the source or candidate target, beyond the effective receptive field of ordinary message passing. Once those bridges are absent, the model is forced to rely on isolated local context. While recent advances in global graph attention~\cite{yun2019GTN,shirzad2023exphormer,VCRTransformer} are designed to capture long-range dependencies, directly applying them to heterogeneous binary graphs is non-trivial due to their scale, sparsity, and mixed code-data semantics.

\noindent
\textbf{\textit{Challenge 2.}} 
Current solutions are also problematically compartmentalized, treating indirect calls, jump tables, tail calls, and returns as separate tasks. This misses useful cross-type structure: return targets depend on preceding calls, and tail calls blur the boundary between call-like and jump-like behavior. Modeling each ICF type independently leaves supervision underutilized and weakens generalization, especially on rare or structurally ambiguous cases.

\vspace{2pt}
\noindent
\textbf{\textit{Challenge 3.}} 
Unlike deterministic algorithms, ML models are highly sensitive to ground-truth quality. Prior studies~\cite{Northcutt2021LabelErrors,Shankar2020ImageNetEval} show that undetected label errors can substantially distort benchmark reliability. However, existing datasets such as Callee~\cite{Callee} and CupidCall~\cite{NeuCall} rely on single-source labels that are inherently prone to either over-approximation or under-approximation, and they lack a clean, standardized test protocol. As a result, it is often unclear whether reported gains reflect real semantic understanding or artifacts of noisy supervision. A clean test set is therefore necessary for credible evaluation.

\vspace{2pt}
\noindent
\textbf{\textit{Challenge 4.}} 
Data leakage remains a pervasive pitfall in learning-based binary analysis~\cite{allamanis2019,zhu2024tygr}. Static linking, library reuse, and code sharing routinely introduce overlap between training and test sets. Without strict partitioning, models can memorize recurring artifacts rather than learn transferable features, inflating reported performance. Our own re-evaluation of the Callee dataset~\cite{Callee} confirms this risk: after enforcing package-level separation and function-level deduplication, the F1 score drops from the reported 94.6\% to 90.92\%. This gap underscores the need for leakage-aware dataset construction and evaluation.


To address \textbf{Challenges 1 and 2}, we present \textit{\mytool{}}, a unified framework for long-range indirect control-flow prediction in stripped binaries. At the architectural level, \mytool{} augments heterogeneous binary graphs with candidate-aware Dual Virtual Hubs: a Global Code Hub (GCH), which provides short routing paths among task-relevant code candidates, and a sparsified Global Data Hub (GDH), which aggregates only a small neighborhood of data nodes around those candidates rather than the entire program. This design preserves long-range reachability while reducing irrelevant global aggregation. On top of this topology, \mytool{} adopts multi-task graph learning to jointly model indirect calls, indirect tail calls, jump tables, and returns, enabling cross-type transfer while retaining task-specific decision boundaries.

To address \textbf{Challenges 3 and 4}, we pair the model with a leakage-aware, noise-controlled evaluation pipeline. We construct a dataset of $15,901$ unique stripped x86\_64 binaries, including $1,351$ binaries with dynamic ground truth, and evaluate under package-level splits, function-level mnemonic-hash deduplication, and a clean test protocol built from verified dynamic positives and absolute negatives. Under this protocol, simply scaling static supervision yields only limited gains. In contrast, the Dual Virtual Hubs improve long-range F1 by up to 9.13 points, multi-task learning adds up to 5.81 points, and the final model outperforms prior baselines by more than 13 F1 on long-range indirect calls.

In a nutshell, we make the following key contributions:

\begin{itemize}
\item \textbf{A unified framework for long-range ICF prediction.} We present \textit{\mytool}, a unified multi-task graph learning framework that leverages Dual Virtual Hubs to jointly predict all indirect control-flow transfers in stripped binaries, effectively resolving long-range dependency and single-task fragmentation issues. 


\item \textbf{A large-scale dataset and rigorous evaluation protocol.}
We construct a dataset of $15,901$ unique stripped x86\_64 binaries, including $1,351$ with dynamic ground truth, and evaluate under package-level splits, function-level mnemonic-hash deduplication, and a clean test protocol built from verified dynamic positives and absolute negatives.

\item \textbf{Strong empirical gains and practical insights.} We show that merely scaling static supervision offers limited benefit, whereas structural repair and cross-type transfer are both important. \mytool\ achieves robust, cross-dataset generalization across both static and dynamic settings, with particularly notable gains on long-range dependencies and structurally complex \textit{inter-procedural} ICF edges. 
\end{itemize}

\begin{table}[h]
\vspace{-3mm}
    \centering
    \caption{ICF types and statistics. The “Share” column reports the proportion of each type among dynamic ICF pairs.}
  \vspace{-2mm}
    \label{tab:cf-types}
    \scalebox{0.9}{ 
    \begin{tabular}{crrrc}
        \toprule
        \textbf{Type} & \textbf{Instr.} & \textbf{Share} & \textbf{Direction} & \textbf{Inter-proc.?} \\
        \midrule
        Indir. Call      & call       & 34.2\% & Forward  & \checkmark \\
        Indir. Tail Call & jmp        & 3.3\%  & Forward  & \checkmark \\
        Jump Table         & jmp        & 15.0\% & Forward  & {\ding{55}} \\
        Return             & ret (pop+jmp)    & 47.5\% & Backward & \checkmark \\
        \bottomrule
    \end{tabular}
    }
\vspace{-5mm}
\end{table}

\section{Background and Related Work}
\label{sec:background}

We begin by motivating our categorization of ICF types, which serves as the foundation for our unified modeling approach. We then survey prior efforts to resolve ICF in binaries, highlighting limitations and key gaps that motivate our work. Finally, we outline the technical foundations underpinning our proposed framework.

\subsection{Indirect Control Flow (ICF) Categorization}
\label{sec:SubtaskPartitioning}

Recovering ICF edges in stripped binaries is challenging due to the loss of high-level semantics and aggressive compiler optimizations, such as bit-test dispatches and merged switch constructs~\cite{Andriesse16,SwitchLowering,pang2020sok}. Unlike rule-based methods, our deep learning approach infers these edges from instruction semantics and structural features. To ensure the model captures subtype-specific behaviors and avoids majority-class bias, we categorize indirect jumps into three distinct subtypes based on their unique control-flow directions and structural patterns: \textit{jump tables}, \textit{indirect tail calls}, and \textit{returns}. 

In contrast, we treat all \textit{indirect calls} as a single unified category. Despite varying dispatch mechanisms, they share a uniform instruction form (i.e., a \texttt{call} to a function entry) and consistent semantics. Our model leverages data nodes and cross-reference edges~\cite{NeuCall} to contextually disambiguate these calls within the graph representation. Table~\ref{tab:cf-types} summarizes the features and distribution of these ICF types, with further rationale provided in Appendix~\ref{sec:appendix-categorization}.

\subsection{Traditional ICF Resolution Techniques}
Traditional methods have long attempted to resolve ICF, but are often limited by rigid assumptions about control-flow structures, compiler behavior, or memory access patterns.

\vspace{2pt}
\noindent
\textbf{\textit{Indirect jump resolution.}} 
Indirect jump resolution has primarily focused on recovering switch-case jump tables. Building on earlier jump-table recovery efforts, subsequent work introduced more sophisticated program-analysis techniques: HRA~\cite{difederico2016jump} combines static and dynamic analysis, while JTR~\cite{cojocar2017jtr} lifts binary code to an intermediate representation and applies SMT-based reasoning. However, these techniques remain limited by the fidelity of binary lifting and the scope of their analyses. The current SOTA static tool, SJA~\cite{nguyen2024}, introduces a forward abstract interpretation framework that avoids pattern matching entirely.
Nevertheless, its evaluation was conducted on LLVM 6.0, which applies simpler jump table lowering, i.e., translating switch statements into predictable indirect jump patterns, than modern compilers that employ more diverse and optimized lowering schemes~\cite{SwitchLowering,Roger08}.
Additionally, SJA's precision degrades at control-flow merge points, and it lacks comprehensive modeling of heap and global memory, which are essential for resolving targets involving indirect memory accesses. Notably, no prior work has explicitly addressed the resolution of indirect tail calls at the binary level, which are more complex as they combine the behavior of indirect calls and returns.

\vspace{2pt}
\noindent
\textbf{\textit{Indirect call resolution.}} 
Resolving indirect calls in stripped binaries remains difficult because source-level type information is unavailable and calling-convention evidence can be incomplete or noisy. Existing solutions make different trade-offs. Signature-based systems such as TypeArmor~\cite{TypeArmor} and TypeSqueezer~\cite{Lin2023TypeSqueezer} rely on recovered type compatibility to prune callees, but their effectiveness depends on the quality of inferred signatures. Pointer-analysis-based systems, including BPA~\cite{BPA} and BinDSA~\cite{Gao2025}, instead reason through memory abstractions and points-to relations. In particular, BinDSA improves precision with field- and context-sensitive analysis, context-sensitive heap reconstruction, and recovered type information, but its effectiveness still depends on accurate lifting, boundary recovery, and memory abstractions, and it explicitly makes precision-soundness trade-offs in difficult cases. Disa~\cite{Wang2025} addresses a complementary prerequisite by improving
function/instruction and memory-block boundary recovery for BPA-style CFG construction, while SchedExec~\cite{Shi2024} uses scheduled runtime execution to eliminate infeasible targets at the cost of scalability. From our perspective, what remains under-exploited across these lines of work is the explicit use of code-data cross-references (xRefs) as first-class representational objects for target prediction. As a result, code regions that interact through the same referenced memory may still remain only weakly coupled during indirect-call reasoning.

\subsection{The Rise of ML-based Single-Task Models}
Machine learning has recently emerged as a powerful alternative, capable of capturing subtle patterns that rule-based techniques often miss. However, existing ML-based approaches are predominantly limited to a single ICF type---indirect calls.


\vspace{2pt}
\noindent
\textbf{\textit{Sequence-/slice-based learning.}} 
Callee~\cite{Callee} and AttnCall~\cite{AttnCall2024} are representative learning-based approaches for indirect-call prediction. Callee applies inter-procedural slicing with expert knowledge, symbolizes the resulting contexts, embeds them with doc2vec, and uses a Siamese network to match callsites and candidate callees; it further transfers knowledge from direct calls to indirect calls via pre-training and fine-tuning~\cite{Callee}. AttnCall replaces the Siamese backbone with a Transformer-style dual-encoder and cross-attention architecture, and introduces an end-to-end instruction representation based on structured instruction vectors and pooling; it also explicitly assumes that semantic matching patterns learned from direct calls transfer to indirect-call prediction~\cite{AttnCall2024}. However, these designs still operate on linearized top/bottom contexts rather than an explicit whole-program graph. Consequently, they do not naturally preserve code-data xRefs as first-class relations, which can limit long-range structural reasoning when evidence is distributed across distant basic blocks, shared data objects, or non-local control/data interactions. Moreover, both methods focus only on indirect calls, leaving the unified modeling of multiple ICF types unexplored.

\vspace{2pt}
\noindent
\textbf{\textit{Graph-based approach.}} 
To address the representational limits of NLP models, CupidCall~\cite{NeuCall} leverages heterogeneous graphs enriched with cross-references (xRefs). However, applying standard Graph Neural Networks (GNNs)~\cite{kipf2017semi} to these structures introduces certain architectural trade-offs. First, the local message passing typical of standard GNNs can blur the distinct semantics of control-flow versus data-flow dependencies. Second, these models are often constrained by limited receptive fields and multi-hop decay~\cite{alon2021on,Topping2022Oversquashing,dwivedi2022LRGB}. Consequently, if static analysis misses a critical xRef, or if the topological distance between nodes exceeds the GNN's effective reach, the localized message passing may become unreliable. Ultimately, while CupidCall significantly advances single-task indirect call resolution, these structural constraints can limit its broader applicability to more complex ICF behaviors.

\begin{figure*} [t]
	\centering
	\includegraphics[width=0.87\textwidth,angle=0,scale=1.0]{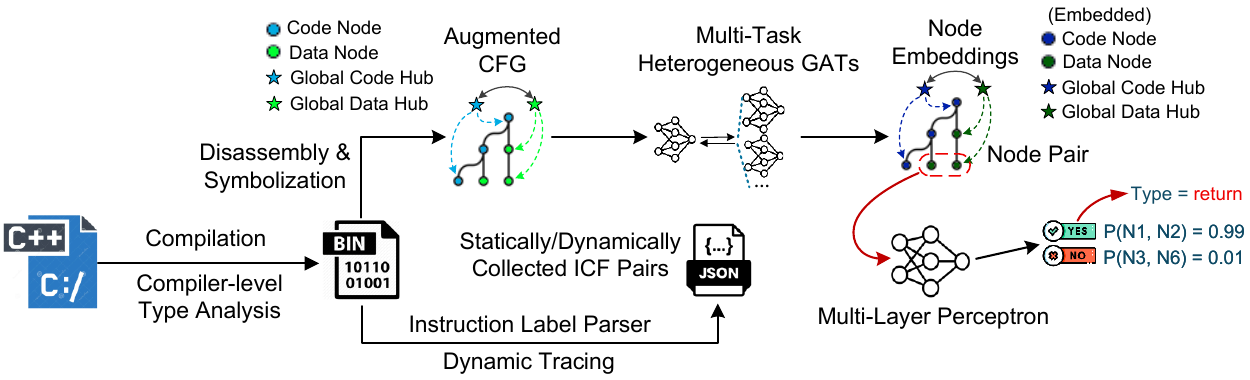}
	\vspace{-3mm}
	\caption{The overview of \mytool's architecture.}
	\label{fig:architecture}
	\vspace{-4mm}
\end{figure*}

\subsection{Graph Representation Learning for Binaries}
Graph-based models are increasingly used in binary analysis because they preserve non-sequential structural relations that are difficult to encode in linearized disassembly. Prior work has shown their usefulness for tasks such as binary provenance analysis, similarity detection, and type inference~\cite{Ji2021Vestige,Qu2023,zhu2024tygr}. These applications naturally involve heterogeneous entities and relations, motivating graph models that can distinguish code from data and control-flow from cross-references~\cite{Kechi22,XiaoWang23,bing2023heterogeneous}.

\vspace{2pt}
\noindent
\textbf{\textit{Heterogeneous graph attention networks.}} 
To model this structure, we construct heterogeneous program graphs whose node and edge types reflect different program entities and relationships. Graph Attention Networks (GATs)~\cite{velivckovic2017graph} are a natural backbone because attention weights can adapt aggregation to noisy and irregular neighborhoods, and prior work has demonstrated the effectiveness of attention-based graph learning on binary code~\cite{Ji2021Vestige,Qu2023}. In heterogeneous settings, type-aware attention mechanisms~\cite{HeterogeneousGAN} further allow message passing to depend on node and edge types, which is important when code semantics and data semantics must remain distinguishable.

\noindent
\textbf{\textit{Long-range reasoning on binary graphs.}} 
Despite these strengths, standard message-passing GNNs still have a limited effective receptive field and often struggle to propagate useful information across many hops~\cite{alon2021on,Topping2022Oversquashing,dwivedi2022LRGB}. This limitation is particularly relevant to binaries, where the evidence needed to resolve an indirect transfer may be distributed across distant basic blocks, shared data objects, and incompletely recovered cross-reference edges. Recent long-range remedies, such as graph transformers and virtual-node style connections~\cite{yun2019GTN,shirzad2023exphormer,VCRTransformer}, are promising, but they are not directly tailored to binary graphs, which are heterogeneous, candidate-sparse, and often only partially recovered. These observations motivate the binary-specific design choices introduced next.

\section{Motivation and Overview}
\label{sec:overview}


We formulate indirect control-flow (ICF) prediction as a \emph{task-aware edge classification} problem on an augmented control-flow graph recovered from a stripped binary. Let
\[
\mathcal{T} = \{\text{ret}, \text{icall}, \text{itcall}, \text{jt}\}
\]
denote the set of ICF types. For each task $\tau \in \mathcal{T}$, we construct task-specific candidate basic-block pairs $(u,v)$ and predict a score $p_{\tau}(u,v)\in[0,1]$ indicating whether an ICF edge of type $\tau$ exists from $u$ to $v$. Collectively, these four tasks recover both the existence and the subtype of indirect control-flow edges. This task-aware formulation is necessary because different ICF types share useful structural regularities, yet they do not share identical source/destination spaces or inductive biases.

\subsection{Key Insight 1: Dual Virtual Hubs}
\label{sec:Dual-Virtual-Hubs}



The central architectural obstacle is not merely graph size, but the mismatch between \emph{where} an ICF transfer is executed and \emph{where} the evidence governing it actually appears. In recovered binaries, the relevant evidence may reside several hops away, may flow through data objects rather than direct control-flow edges, or may be only partially exposed because code-data xRefs are incompletely recovered. Consequently, even graph-based models such as CupidCall~\cite{NeuCall} can degrade when ordinary message passing must carry semantic information across distant or fragmented structures.

To address this, \mytool\ introduces \emph{candidate-aware Dual Virtual Hubs} inspired by global graph attention~\cite{yun2019GTN,shirzad2023exphormer,VCRTransformer}: a \textit{Global Code Hub} (GCH) and a \textit{Global Data Hub} (GDH). At a high level, the GCH provides sparse fallback routes among task-relevant code candidates, while the GDH aggregates data-side evidence associated with those candidates rather than indiscriminately mixing the entire program. By synchronizing these two hubs, \mytool\ creates short cross-domain communication paths between distant code and data evidence. This design alleviates long-range bottlenecks while preserving the sparse, heterogeneous structure of binary graphs.

\subsection{Key Insight 2: Multi-Task Learning}
\label{sec:Multi-Task}


A second obstacle is task fragmentation. Existing models~\cite{Callee,AttnCall2024,nguyen2024,NeuCall} typically address only one ICF subtype at a time, even though the underlying behaviors are related. Return targets depend on earlier calls, indirect tail calls combine call-like and jump-like behavior, and the same recovered graph may contain evidence that is useful across multiple ICF types. Training separate models therefore underutilizes supervision and is especially brittle for rare or structurally ambiguous cases.

This motivates our use of Multi-Task Learning (MTL)~\cite{evgeniou2004regularized,parameswaran2010large,evgeniou2005learning}. \mytool\ employs a shared representation to capture reusable structural patterns across ICF types, while preserving task-specific decision boundaries through separate prediction heads. This arrangement enables \emph{inductive transfer}: abundant tasks (e.g., returns) help regularize and guide the learning of rarer tasks (e.g., indirect tail calls), improving generalization without collapsing all ICF types into a single homogeneous objective.

\subsection{\mytool\ Workflow}
\label{sec:workflow}

Figure~\ref{fig:architecture} summarizes both the \emph{offline supervision pipeline} used for training/evaluation and the \emph{model pipeline} used at inference. To avoid ambiguity, only Stage~I is offline; Stages~II--IV constitute the deployable prediction path.


\vspace*{2pt}
\noindent\textbf{\textit{I. Offline supervision pipeline ($\S\ref{sec:data-collection}$).}}
For training and evaluation only, we collect two complementary supervision sources: compiler-derived static ICF pairs obtained from a custom LLVM toolchain, and dynamic ICF traces collected with Intel Pin~\cite{intel-pin}. These labels are never used as runtime input; they serve solely as supervision and evaluation signals.


\vspace*{2pt}
\noindent\textbf{\textit{II. Augmented CFG construction ($\S\ref{sec:cfg-construction}$).}}
Given a stripped binary, we use \texttt{angr}~\cite{Yan16} to recover a base control-flow graph and code-data xRefs via \emph{symbolization}, which interprets immediate values and memory references as potential references between code and data~\cite{wang2015reassembleable}. We then augment this recovered graph with candidate-aware Dual Virtual Hubs so that task-relevant code candidates and nearby data evidence can interact through sparse long-range routes even when explicit xRef chains are incomplete.


\vspace*{2pt}
\noindent\textbf{\textit{III. Heterogeneous graph encoding and edge scoring ($\S\ref{sec:base-hgat}$).}}
A Heterogeneous Graph Attention Network (HGAT) encodes the augmented graph into relation-aware node representations. For each task-specific candidate pair, \mytool\ combines the source and destination embeddings and predicts the probability that the corresponding ICF edge exists. To study the effect of supervision quality without changing the architecture, we train the same model under separate static and dynamic supervision regimes.


\begin{figure}[t]
    \centering
    \includegraphics[width=0.45\textwidth]{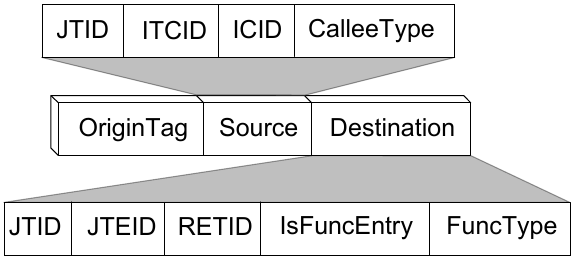}
    \vspace{-2mm}
    \caption{Instruction label.}
    \label{fig:instrlabel}
    \vspace{1mm}
    {\footnotesize
    \begin{tabular}{ll}
        OriginTag: & Identifier of the source file. \\
        JTID: & Identifier of the jump table. \\
        ITCID: & Identifier of the indirect tail call. \\
        ICID: & Identifier of the indirect call. \\
        CalleeType: & Callee type of the \texttt{call} instruction. \\
        JTID-JTEID: & Entry index for the jump table ID. \\
        RetID: & Return site ID (for functions with multiple \texttt{ret}s). \\
        isFuncEntry: & 1 = Function entry; 0 = Not an entry. \\
        FuncType: & Type identifier for the current function. \\
    \end{tabular}
    }
    \vspace{-5mm}
\end{figure}

\vspace*{2pt}
\noindent\textbf{\textit{IV. Task-aware multi-task learning ($\S\ref{sec:multitask-hgat}$).}}
On top of the shared HGAT encoder, \mytool\ adds task-specific graph heads and classifiers for ICF variants. This design allows the model to share structural knowledge across tasks while preserving subtype-specific prediction boundaries. During inference, the same statically recoverable graph construction path is used, and the task-aware heads score candidate pairs across all four ICF types.

\vspace{-2mm}
\section{System Design}
\label{sec:methodology}



This section instantiates the four stages summarized in Figure~\ref{fig:architecture}: offline supervision collection, augmented CFG construction, heterogeneous graph encoding, and task-aware multi-task learning.

\subsection{Clean Evaluation and Source Separation}
\label{sec:data-collection}


\noindent
\textbf{\textit{Static ICF pairs}.} \label{sec:static icf collection}
We customize the LLVM frontend and backend to collect ground-truth ICF pairs during compilation. For indirect calls and indirect tail calls, we use LLVM-CFI~\cite{LLVM-CFI} annotations and address-taking criteria, following a type-based strategy inspired by Modular CFI~\cite{NiuTan2014MCFI}. For jump tables, we intercept instruction selection to recover dispatch-target mappings. Return edges are reconstructed by pairing callsites with their fall-through successors while accounting for tail-call optimization. All collected pairs are indexed by our instruction-labeling system and later mapped back to final binary addresses.
We provide more details in Appendix~\ref{sec:appendix-static-icf}.

\vspace{2pt}
\noindent
\textbf{\textit{Instruction-level labeling.}} \label{sec:instruction-label} 
\label{para:inst_labeling}
An instruction may serve as a source, destination, or both across multiple ICF transfers. To represent this, we design a compact, delimiter-separated labeling format that encodes both source and destination metadata into a single string (see Figure~\ref{fig:instrlabel}). The source segment records outgoing edge attributes (e.g., dispatch context), while the destination segment captures incoming properties (e.g., target type). This compositional structure allows one instruction to support multiple ICF types and incremental updates. Post-generation, \mytool\ parses these labels to reconstruct ICF pairs and lifts them to the basic block level.


\vspace{2pt}
\noindent
\textbf{\textit{Dynamic ICF pairs.}} \label{sec:dynamic-collection}
We complement static supervision with dynamic traces collected by a customized Intel Pin tool~\cite{intel-pin}. Tracing is triggered by each project's native test suite and supplementary tests when available. Dynamic indirect jumps are post-classified into subtypes by matching dispatch sites against known tail-call and jump-table locations. All traced events are then lifted to basic blocks using the same mapping pipeline as our static representation.

\vspace{2pt}
\noindent
\textbf{\textit{Clean test protocol.}} \label{sec:clean-test}
Static and dynamic supervision have complementary failure modes: static analysis over-approximates, whereas dynamic tracing is incomplete. We therefore evaluate with a clean test protocol that treats dynamically observed targets as positives and samples an equal number of negatives only from the region outside the static candidate set. This avoids penalizing dynamically unseen but statically plausible edges. 
For example, if an indirect callsite has 3 dynamic targets and 10 static candidates among 1,000 total functions, then the 3 dynamic edges are treated as positives, and 3 negatives are sampled only from the remaining $1{,}000-10=990$ non-candidate functions. In our dataset, this absolute-negative region covers 98.7\% of the inter-procedural control-flow space. \emph{For jump tables, whose compiler-derived labels are highly reliable, we later use the static labels as ground truth rather than dynamic positives.}

\vspace{2pt}
\noindent\textbf{\textit{Supervision source separation.}} 
Static and dynamic supervision also provide different trade-offs: the former offers broader coverage, while the latter offers higher precision. Unlike prior work that relies on a single source~\cite{Callee,NeuCall}, we treat compiler-derived and dynamically traced ICF pairs as two independent supervision sources. We therefore train two architecturally identical models, one under static supervision and the other under dynamic supervision, and evaluate both on the same \hyperref[sec:clean-test]{clean test set}. This design isolates the effect of supervision quality from architectural differences and enables a fair comparison between comprehensive-but-noisy and precise-but-incomplete supervision regimes.

\subsection{Augmented CFG Construction }
\label{sec:cfg-construction}

\noindent \textbf{\textit{Base graph construction.}} 
Given a stripped binary, we recover a heterogeneous augmented CFG (ACFG) using \texttt{angr}~\cite{Yan16} and symbolization~\cite{wang2015reassembleable}. Symbolization interprets immediate values and memory references as potential code-data xRefs. Compared with prior ACFG formulations~\cite{NeuCall}, we additionally preserve deterministic direct jumps as explicit edge types because they expose the intra-procedural skeleton needed for jump-table and return reasoning. To maintain strict feature/label separation, all target ICF edges are excluded from the input graph and used only as supervision.

\vspace{2pt}
\noindent
\textbf{\textit{Dual global hubs.}} 
Ideally, an explicit local cross-reference path (e.g., $\texttt{src}\rightarrow \texttt{data}\rightarrow \texttt{dst}$) provides the strongest structural evidence for ICF prediction~\cite{NeuCall}. In recovered binary graphs, however, such paths are often incomplete: the exact shared data object may be missing, only part of the chain may be recovered, or the relevant evidence may be fragmented across nearby wrappers and basic blocks. Under ordinary message passing, these breaks either sever connectivity or force useful evidence to travel through many local hops, where it weakens.
To mitigate this bottleneck, we augment the graph with two virtual nodes: a \emph{Global Code Hub} (GCH, $v_{\mathrm{GCH}}$) and a \emph{Global Data Hub} (GDH, $v_{\mathrm{GDH}}$). The GCH provides a sparse code-side fallback among task-relevant code candidates, while the GDH aggregates nearby data evidence associated with those candidates. A bidirectional link between the two hubs allows control-side and data-side summaries to interact within constant depth. We call them ``global'' because they enable long-range communication across the graph, not because they connect densely to all nodes.

\begin{table}[t]
\centering
\caption{Taxonomy of ICF transfers and their candidates.}
\vspace{-2mm}
\label{tab:indirect_types}
\scalebox{0.82}{%
\begin{tabular}{@{}lll@{}}
\toprule
\textbf{Type ($\tau$)} & \textbf{Potential Source ($C_{src}^{\tau}$)} & \textbf{Potential Destination ($C_{dst}^{\tau}$)} \\ \midrule
\textbf{Indir. Call }       & Indir. call sites       & Function entry blocks \\
\textbf{Indir. Tail Call } & Indir. jump sites       & Function entry blocks \\
\textbf{Jump Table }         & Indir. jump sites       & Intra-function basic blocks \\
\textbf{Return}              & Return basic blocks       & Post-callsite basic blocks \\ \bottomrule
\end{tabular}%
}
\vspace{-5mm}
\end{table}

\vspace{2pt}
\noindent
\textbf{\textit{Candidate-aware pruning.}} \label{sec:candidate-aware}
Binary graphs are large, but ICF prediction is candidate-sparse. Standard global attention connects all node pairs and incurs $\mathcal{O}(N^2)$ cost; even restricting attention to one task would still require $\mathcal{O}(|C^{\tau}_{src}|\cdot |C^{\tau}_{dst}|)$ direct routing. 

We therefore connect only task-relevant candidates through the hubs, reducing the additional routing footprint to a sparse, near-linear form.
Formally, for each task $\tau \in \mathcal{T}=\{\texttt{ret},\texttt{icall},\texttt{itcall},\texttt{jt}\}$, we define a source candidate set $C^{\tau}_{src}$ and a destination candidate set $C^{\tau}_{dst}$ according to Table~\ref{tab:indirect_types}. The GCH connects only to nodes in these two code candidate sets. The GDH is also sparsified: instead of connecting to all data nodes, it connects only to the task-specific data neighborhood
\[
D_{\tau}=\{d\in V_d \mid \mathrm{dist}_{xref}(d, C^{\tau}_{src}\cup C^{\tau}_{dst}) \le 2\},
\]
where
\[
\mathrm{dist}_{xref}(d, C)=\min_{u\in C}\mathrm{dist}_{G_{xref}}(d,u),
\]
and $G_{xref}$ is the subgraph induced by explicit code-data relations in the recovered base graph. This definition preserves nearby data evidence that is often useful for ICF resolution while filtering out unrelated global data nodes. We select the 2-hop radius empirically after evaluating 1, 2, and 3 hops (detailed in Appendix~\ref{appendix:GDH Pruning}).

A further requirement is that the same graph-construction procedure must be available during inference. Hence, all candidate sets must be derivable purely from static frontend outputs. This creates an ambiguity for indirect jumps: as shown in Table~\ref{tab:indirect_types}, jump tables and indirect tail calls may originate from similar indirect-jump sites but have different destination spaces. Rather than forcing a brittle frontend subtype decision, we disable static indirect-jump resolution and conservatively route unresolved indirect-jump blocks to the GCH as generic source candidates. This slightly increases hub edges, but preserves source-side recall and leaves subtype disambiguation to the downstream model. Consequently, the statically constructed sets $C^{\tau}_{src}$, $C^{\tau}_{dst}$, and $D_{\tau}$ serve only as structural priors for graph construction, not as supervision labels.

\vspace*{2pt}
\noindent\textbf{\textit{Node initial embedding.}} 
We define initial node representations tailored to their specific roles. Ordinary code nodes 
are initialized with PalmTree embeddings~\cite{li2021palmtree} to capture fine-grained instruction semantics. For data nodes, we advance beyond CupidCall's bare normalized addresses~\cite{NeuCall} by concatenating them with lightweight structural features. Finally, the two newly proposed hub nodes (GCH \& GDH) use learnable type embeddings rather than binary-derived features to capture their overarching semantic roles.

\subsection{Heterogeneous Graph Attention Networks}
\label{sec:base-hgat}

After constructing the augmented graph in Section~\ref{sec:cfg-construction}, we encode it with a Heterogeneous Graph Attention Network (HGAT)~\cite{HeterogeneousGAN}. Compared with homogeneous GNNs, HGAT preserves the semantics of different relation types, which is important in our setting because control-flow edges and code-data xRefs play distinct roles in ICF reasoning.

\vspace{2pt}
\noindent \textbf{\textit{Base graph formulation.}} 
We represent each recovered program as a heterogeneous graph $G=(V,E,R,T)$,
where $V=V_c\cup V_d$ contains code nodes and data nodes, $R$ is the relation-type set, $T=\{\texttt{code},\texttt{data}\}$ is the node-type set, and $E\subseteq V\times R\times V$ contains typed control-flow and xRef relations. Following prior ACFG-style formulations~\cite{NeuCall}, we preserve direct control-flow edges and explicit cross-references, and we optionally add reverse relations for bidirectional message passing. To avoid leakage from frontend heuristics, we disable \texttt{angr}'s~\cite{Yan16} indirect-jump resolver, so no heuristically recovered indirect targets appear in the input graph.

\vspace{2pt}
\noindent
\textbf{\textit{Relation-aware HGAT encoder.}}
Let $\mathcal{N}_{v}^{r}=\{u\mid (u,r,v)\in E\}$ denote the neighbors of node $v$ under relation $r$. For each relation $r$, target node $v$, and layer $l$, the $K$-head attention coefficients and relation-specific aggregated message are
\begin{equation}
\alpha_{uv}^{(r,k)}
=
\mathrm{softmax}_{u\in\mathcal{N}_{v}^{r}}
\left(
\mathrm{LeakyReLU}
\left(
(\mathbf{a}_{r}^{(k)})^{\top}
\big[
\mathbf{W}_{r}^{(k)}\mathbf{h}_{u}^{(l)}
\parallel
\mathbf{W}_{r}^{(k)}\mathbf{h}_{v}^{(l)}
\big]
\right)
\right),
\end{equation}
\begin{equation}
\mathbf{m}_{v}^{(r,l)}
=
\frac{1}{K}
\sum_{k=1}^{K}
\sum_{u\in\mathcal{N}_{v}^{r}}
\alpha_{uv}^{(r,k)}\mathbf{W}_{r}^{(k)}\mathbf{h}_{u}^{(l)},
\end{equation}
where $\mathbf{h}_{v}^{(l)}$ is the hidden state of node $v$, $\parallel$ denotes concatenation, and $\mathbf{W}_{r}^{(k)}$ and $\mathbf{a}_{r}^{(k)}$ are relation-specific learnable weights. These base messages $\mathbf{m}_v^{(r,l)}$ elegantly encapsulate localized, relation-aware context prior to integration with global routing signals.

\vspace{2pt}
\noindent
\textbf{\textit{Extended graph definition.}}
For a single task $\tau$, we extend $G$ to
\[
G_{\tau}^{+}=(V_{\tau}^{+},E_{\tau}^{+},R_{\tau}^{+},T_{\tau}^{+}),
\]
where
\[
V_{\tau}^{+}=V\cup\{v_{\mathrm{GCH}},v_{\mathrm{GDH}}\},
\qquad
T_{\tau}^{+}=T\cup\{\texttt{gcode},\texttt{gdata}\},
\]
and
\[
\begin{aligned}
R_{\tau}^{+}=R\cup \{&
r_{src2gch},r_{gch2src},
r_{dst2gch},r_{gch2dst},\\
&
r_{data2gdh},r_{gdh2data},
r_{gch2gdh},r_{gdh2gch}\}.
\end{aligned}
\]
Using the candidate sets $C^{\tau}_{src}$, $C^{\tau}_{dst}$, and $D_{\tau}$ defined in Section~\ref{sec:cfg-construction}, we add the hub edges
\[
\begin{aligned}
E_{\tau}^{hub}
={}&
\{(u,r_{src2gch},v_{\mathrm{GCH}})\mid u\in C^{\tau}_{src}\}
\cup
\{(v_{\mathrm{GCH}},r_{gch2src},u)\mid u\in C^{\tau}_{src}\} \\
&\cup
\{(u,r_{dst2gch},v_{\mathrm{GCH}})\mid u\in C^{\tau}_{dst}\}
\cup
\{(v_{\mathrm{GCH}},r_{gch2dst},u)\mid u\in C^{\tau}_{dst}\} \\
&\cup
\{(u,r_{data2gdh},v_{\mathrm{GDH}})\mid u\in D_{\tau}\}
\cup
\{(v_{\mathrm{GDH}},r_{gdh2data},u)\mid u\in D_{\tau}\} \\
&\cup
\{(v_{\mathrm{GCH}},r_{gch2gdh},v_{\mathrm{GDH}}),
(v_{\mathrm{GDH}},r_{gdh2gch},v_{\mathrm{GCH}})\}.
\end{aligned}
\]
The final edge set is
\[
E_{\tau}^{+}=E\cup E_{\tau}^{hub}.
\]


\vspace{2pt}
\noindent
\textbf{\textit{Routing intuition.}}
These hub edges implement a gather-and-broadcast mechanism. Source and destination candidates send task-relevant code semantics to the GCH, while data nodes in $D_{\tau}$ provide nearby data evidence to the GDH. The bidirectional GCH$\leftrightarrow$GDH link then exchanges control-side and data-side summaries. When a local $\texttt{src}\rightarrow\texttt{data}\rightarrow\texttt{dst}$ chain is missing or fragmented, the model can still exploit two complementary fallback routes: a code-side route through the GCH, and a data-side route through $\{\texttt{src},\texttt{dst}\}\rightarrow \mathrm{GCH}\rightarrow \mathrm{GDH}\rightarrow D_{\tau}$.

\vspace{2pt}
\noindent
\textbf{\textit{Hub-gated fusion.}}
Let $\phi:V_{\tau}^{+}\rightarrow T_{\tau}^{+}$ denote the node-type mapping. For each ordinary node $v$, we partition its incident relations into local relations $R_{v}^{loc}$ and hub relations $R_{v}^{hub}$, and aggregate them separately:
\[
\mathbf{m}_{v,loc}^{(l)}=\sum_{r\in R_{v}^{loc}}\mathbf{m}_{v}^{(r,l)},
\qquad
\mathbf{m}_{v,hub}^{(l)}=\sum_{r\in R_{v}^{hub}}\mathbf{m}_{v}^{(r,l)}.
\]
To prevent global routing from overwhelming strong local evidence, we compute a \emph{scalar} local-conditioned gate
\begin{equation}
g_{v}^{(l)}
=
\sigma\!\left(
(\mathbf{w}_{\phi(v)})^{\top}
\big[
\mathbf{m}_{v,loc}^{(l)}\parallel \mathbf{h}_{v}^{(l)}
\big]
+
b_{\phi(v)}
\right),
\end{equation}
where $g_{v}^{(l)}\in(0,1)$ and $\sigma$ is the sigmoid function. The updated node state is
\begin{equation}
\mathbf{h}_{v}^{(l+1)}
=
f\!\left(
\mathrm{LN}_{\phi(v)}
\big(
\mathbf{m}_{v,loc}^{(l)}
+
\alpha_{\phi(v)} g_{v}^{(l)} \mathbf{m}_{v,hub}^{(l)}
+
\mathbf{W}_{\phi(v)}^{res}\mathbf{h}_{v}^{(l)}
\big)
\right),
\end{equation}
where $f$ is a nonlinearity (we use ELU), $\alpha_{\phi(v)}$ is a learnable scalar, and $\mathrm{LN}_{\phi(v)}$ is a node-type-specific LayerNorm. Only ordinary nodes use this gated fusion; hub nodes follow the standard HGAT update.

\vspace{2pt}
\noindent
\textbf{\textit{Mitigating hub-side bottlenecks.}} 
To prevent information bottlenecks, we mitigate hub-side oversquashing via three structural constraints. 
First, \emph{candidate-aware sparse connectivity} restricts links exclusively to task-specific nodes, minimizing noise influx. Second, \emph{relation-specific multi-head attention} preserves distinct semantic subspaces for diverse interactions (e.g., src-to-hub vs. data-to-hub) rather than collapsing them. Finally, \emph{hub-gated fusion} selectively injects global context, ensuring hubs act as routing shortcuts that complement, rather than overwhelm, local ACFG evidence.  Together, these mechanisms support efficient long-range communication without reducing hubs to dense, undifferentiated sinks.



\vspace{2pt}
\noindent
\textbf{\textit{Task-specific edge scoring.}}
The extended HGAT model is trained end-to-end using a binary edge classification objective.
For each task $\tau$, let
\[
\mathcal{P}_{\tau}\subseteq C^{\tau}_{src}\times C^{\tau}_{dst}
\]
denote the candidate source--destination pairs to be scored on $G_{\tau}^{+}$. For each $(u,v)\in\mathcal{P}_{\tau}$, we concatenate the source and
destination representations and apply a task-specific two-layer MLP:
\begin{equation}
\begin{aligned}
\mathbf{r}_{uv}^{\tau}
&=
\operatorname{GELU}\!\left(
\mathbf{W}_{\tau,1}
\left[
\mathbf{h}_{u}^{(L)}
\parallel
\mathbf{h}_{v}^{(L)}
\right]
+
\mathbf{b}_{\tau,1}
\right), \\
\hat{y}_{uv}^{\tau}
&=
\sigma\!\left(
\mathbf{w}_{\tau,2}^{\top}
\mathbf{r}_{uv}^{\tau}
+
b_{\tau,2}
\right),
\end{aligned}
\label{eq:edge_prediction}
\end{equation}
where $\mathbf{W}_{\tau,1}\in\mathbb{R}^{256\times512}$,
$\mathbf{b}_{\tau,1}\in\mathbb{R}^{256}$,
$\mathbf{w}_{\tau,2}\in\mathbb{R}^{256}$, and
$\mathbf{r}_{uv}^{\tau}\in\mathbb{R}^{256}$.
Each task $\tau$ uses an independent endpoint-scoring MLP.
We optimize the weighted binary cross-entropy loss
\begin{equation}
\mathcal{L}_{\tau}^{\mathrm{BCE}}
=
-
\sum_{(u,v)\in\mathcal{P}_{\tau}}
\left[
w_{\mathrm{pos}}^{\tau}
y_{uv}^{\tau}
\log \hat{y}_{uv}^{\tau}
+
\left(1-y_{uv}^{\tau}\right)
\log\left(1-\hat{y}_{uv}^{\tau}\right)
\right],
\label{eq:weighted_bce}
\end{equation}
where $y_{uv}^{\tau}\in\{0,1\}$ is the label for task $\tau$, and
$w_{\mathrm{pos}}^{\tau}$ is the task-specific positive weight computed
from the training split.


\subsection{Multi-Task Heterogeneous GATs}
\label{sec:multitask-hgat}

\begin{figure} [t]
	\centering
	\includegraphics[width=0.47\textwidth,angle=0,scale=1.0]{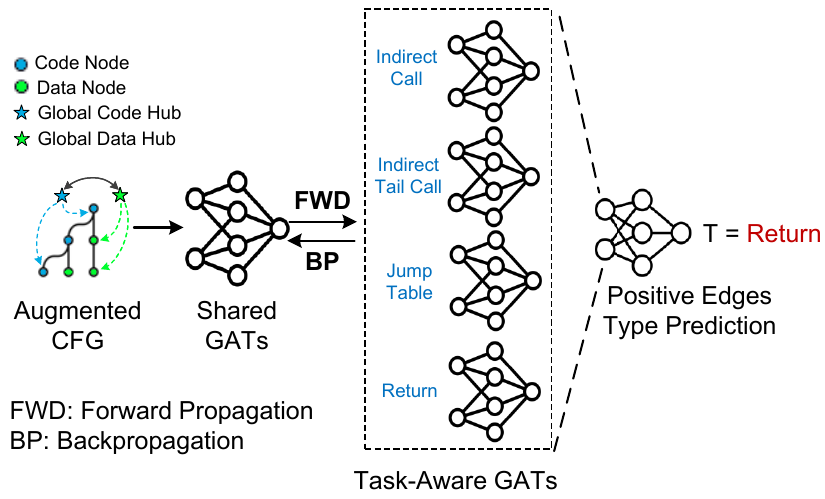}
	\vspace{-3mm}
	\caption{The overview of multi-task learning architecture.}
	\label{fig:MTL}
	\vspace{-5mm}
\end{figure}

To jointly predict all four ICF types, we extend the HGAT into a multi-task learning (MTL) architecture, as depicted in Figure~\ref{fig:MTL}. A shared heterogeneous encoder (comprising the computationally intensive lower layers) feeds four lightweight task-specific GATs, each driving a dedicated binary edge classifier. To strictly prevent label contamination, each classifier trains exclusively on its own positive edges and independently sampled negatives. This isolation preserves task-specific nuances while enabling reciprocal cross-type transfer, substantially outperforming independent single-task models. Additionally, an auxiliary edge-type classification head---trained solely on positive edges---forces the shared encoder to explicitly separate semantically similar behaviors, yielding highly type-discriminative representations.

\vspace{2pt}
\noindent \textbf{\textit{Unified MTL graph.}} 
The MTL framework aggregates all task candidates into a unified graph $G_{MTL}^{+}$. 
To avoid brittle frontend subtype decisions and keep graph construction inference-available, we uniformly route all ambiguous indirect-jump \textit{sources} (tail-calls vs. jump tables) to the GCH. Crucially, this is paired with an asymmetric routing design where jump-table \textit{targets} (intra-procedural blocks) are strictly excluded. We enforce this isolation to prevent injecting intra-procedural noise into the global hub---an issue we empirically observed to degrade jump-table prediction in single-task settings. As a result, the GCH explicitly avoids absorbing local target semantics, preserving its integrity as a predominantly inter-procedural anchor. Ultimately, this asymmetric topology provides an elegant structural prior for the MTL framework: it naturally restricts the task-specific jump-table head to rely on the local ACFG topology, while enabling the tail-call head to exploit the GCH for global reachability, thereby maintaining precision across all scales.

\vspace{2pt}
\noindent
\textbf{\textit{Shared encoder without label leakage.}} 
Our shared encoder serves solely to extract common topological features, while task-specific GATs train exclusively on their respective labels. This strict isolation prevents \emph{label leakage}~\cite{li2022label,Liu2024LabelLI}---a critical flaw where ground-truth from one task inadvertently influences another. By fully decoupling the supervision signals, our architecture exploits the generalization benefits of multi-task learning while maintaining the rigorous evaluation bounds of independent single-task models.

\vspace{2pt}
\noindent
\textbf{\textit{Overall loss.}}
We optimize
\begin{equation}
\mathcal{L}_{total}
=
\sum_{\tau\in\mathcal{T}}\lambda_{\tau}\mathcal{L}^{\mathrm{BCE}}_{\tau}
+
\lambda_{type}\mathcal{L}^{\mathrm{CE}}_{type},
\end{equation}
where $\mathcal{L}^{\mathrm{CE}}_{type}$ is an optional auxiliary type-classification loss computed only on positive training pairs to encourage subtype-discriminative shared representations. In our experiments, we set $\lambda_{\tau}=1$ for all tasks and $\lambda_{type}=1$ when the auxiliary head is enabled. 
This formulation encourages the model to make consistent and accurate predictions across tasks while leveraging auxiliary supervision to strengthen learning during training.



\section{Evaluation}
\label{sec:evaluation}


Prior work~\cite{Callee,NeuCall} has shown that more accurate ICF recovery strengthens downstream security analyses such as binary diffing and hybrid fuzzing by providing a more faithful structural backbone. In this paper, however, our goal is to evaluate that foundational layer directly. Downstream applications introduce many additional components and confounding factors, which can obscure whether an observed gain truly comes from improved ICF recovery. We therefore focus our evaluation on \mytool's core predictive capability itself: recovering high-fidelity ICF edges across all major ICF types---not merely indirect calls---in stripped binaries. By delivering high-fidelity predictions,
\mytool\ provides the comprehensive control-flow recovery essential for robust downstream defenses.


This section proceeds as follows. We first describe the construction of our training dataset, which is designed to provide reliable supervision across all four ICF types. We then present the experimental setup, including strict package-level splitting, function-level deduplication, and hyperparameter tuning. The evaluation itself is organized along three main axes: \emph{supervision source} (Dynamic vs.\ Static vs.\ Large Static), \emph{structural difficulty} (Overall vs.\ Long-range), and \emph{training paradigm} (Single-task vs.\ Multi-task). Concretely, we first evaluate the full single-task model with Dual Hubs under different supervision regimes to determine the strongest training setting (RQ1). We then isolate the contribution of the Dual Virtual Hubs, focusing on long-range and otherwise hard-to-reach pairs where local message passing is most challenged (RQ2). Next, we introduce multi-task learning (MTL) to measure the additional gains from cross-type transfer under the same setting (RQ3). We subsequently compare the resulting system with prior state-of-the-art (SOTA) approaches on both established benchmarks and our curated dataset (RQ4). Finally, we assess the practical efficiency and scalability of the MTL framework and Dual Virtual Hubs (RQ5).










\begin{itemize}
\item \textbf{RQ1} \textit{(Single-task performance with Dual Hubs):} Under the same clean-test protocol, how does our single-task model with Dual Hubs perform under dynamic, static, and large-scale static supervision?

\item \textbf{RQ2} \textit{(Long-range ICF pairs analysis):} How critical are the Dual Virtual Hubs for resolving long-range and otherwise unreachable ICF pairs?

\item \textbf{RQ3} \textit{(Multi-task model performance):} How does the multi-task model compare with the single-task model under the same supervision setting?

\item \textbf{RQ4} \textit{(Comparative evaluation):} How does our model compare with existing ICF prediction approaches on both prior benchmarks and our comprehensive curated dataset?

\item \textbf{RQ5} \textit{(Efficiency and scalability):} What are the computational costs and scalability characteristics of our MTL framework and Dual Virtual Hubs?
\end{itemize}

\begin{table}[t]
    \centering
    \caption{The statistics of our collected ICF edges. $^{(s)}$ = static (compiler-level), $^{(d)}$ = dynamic.}
   \vspace{-2mm}
    \label{tab:edge-stats-compressed}
    \small
    \resizebox{0.41\textwidth}{!}{%
    \begin{tabular}{l|rrr}
        \hline
        \textbf{Type} & \textbf{\# of Bin.} & \textbf{\# of Sites} & \textbf{\# of Pairs} \\
        \hline
        Indir. Call$^{(s)}$        & 12.6k  & 2.6M   & 63.4M  \\
        Indir. Tail Call$^{(s)}$   & 2.1k   & 42.3k  & 2.9M   \\
        Jump Table$^{(s)}$         & 13.3k  & 325.8k & 2.6M   \\
        Return$^{(s)}$             & 12.7k  & 6.3M   & 180.1M \\
        \hline
        Indir. Call$^{(d)}$        & 858  & 58.1k & 92.1k  \\
        Indir. Tail Call$^{(d)}$   & 209  & 3.1k  & 8.8k   \\
        Jump Table$^{(d)}$         & 1171 & 15.1k & 40.3k  \\
        Return$^{(d)}$             & 928  & 68.6k & 128.1k \\
        \hline
    \end{tabular} }
\vspace{-5mm}
\end{table}

\subsection{Training Dataset Construction}
\label{sec:dataset}



To train and evaluate \mytool\ consistently, we require a unified corpus with reliable ground truth covering all four indirect control-flow (ICF) types. Existing datasets are fragmented and each captures only part of this space. SJA~\cite{nguyen2024} focuses on jump tables collected with an older LLVM toolchain; CupidCall~\cite{NeuCall} is trained exclusively on statically inferred indirect-call pairs; and Callee~\cite{Callee} relies solely on dynamically traced indirect-call pairs. Such single-source supervision schemes limit diversity, complicate fair comparison, and prevent comprehensive multi-task training across ICF types. We therefore construct a new large-scale dataset explicitly designed to provide consistent supervision and reliable ground truth for all four ICF transfers.

Our dataset is built from 12,972 C/C++ projects in the Arch Linux ecosystem. Each package is compiled from source using a custom LLVM-based toolchain (v19.1.5) integrated with Arch's \texttt{makepkg} and \texttt{PKGBUILD} infrastructure. After deduplication, we retain only binaries containing at least one static ICF pair, extracted via our instruction-labeling parser. For packages with native test suites, we additionally perform dynamic tracing to obtain precise runtime ICF pairs. To reduce the coverage bias of native tests on more complex software, we further select 20 ubiquitous medium-sized real-world projects (e.g., PostgreSQL and Wireshark) and use LLM-generated supplementary tests to trigger deeper execution paths. This augmentation substantially enriches the dynamic ground truth for complex applications. Appendix~\ref{appendix:dataset} provides the detailed dataset construction pipeline.

Ultimately, this pipeline yields $15,901$ unique stripped binaries, including 1,351 with dynamic ground truth---substantially larger than prior dynamic resources such as Callee~\cite{Callee}. 
Table~\ref{tab:edge-stats-compressed} summarizes the resulting edge-level statistics. As expected, indirect tail calls remain much rarer than indirect calls and returns, reflecting the fact that tail-call optimization is governed by strict optimization and Application Binary Interface (ABI) constraints~\cite{minamide2003selective}.

\subsection{Experimental Setup} \label{sec:setup}

All experiments were conducted on a server equipped with an Intel Xeon 8358 CPU, 512GB RAM, and four 80GB NVIDIA A100 GPUs. Our implementation relies on PyTorch 2.3, Deep Graph Library (DGL) 2.4, and angr 9.2.157.

\noindent

\vspace*{2pt}
\noindent\textbf{\textit{Package-level stratified data splitting.}}
To prevent data leakage and ensure fair evaluation, we adopt a package-level multi-label proportional sampling strategy~\cite{MultiLabel}. Because binaries within the same package often share code templates, compiler options, and library dependencies, we assign entire packages to a single split. This design preserves structural independence across datasets while maintaining a balanced distribution of ICF edge types. Please refer to Appendix~\ref{appendix:data-split} for more details.

\vspace*{2pt}
\noindent\textbf{\textit{Noise-aware training, noise-free evaluation.}} 
We deliberately separate the rules for training from those for evaluation. During training, we tolerate some label noise to preserve coverage; during evaluation, we use only low-noise labels. For jump tables, compiler-derived labels are sufficiently reliable and therefore serve as the ground truth for all models. For indirect calls, indirect tail calls, and returns, verified dynamic targets define the true edges. When training with dynamic supervision, however, tracing incompleteness makes it unsafe to treat every unobserved candidate as a negative. We therefore depart from Callee~\cite{Callee}: if an unobserved candidate remains statically type-consistent, we discard it from the negative pool rather than labeling it negative, thereby reducing false negatives ($FN$) introduced by incomplete execution coverage. In contrast, when training with static supervision, we keep the statically collected positives---including unavoidable false positives---to test robustness under broader but noisier supervision. Evaluation remains strictly noise-free: for inter-procedural tasks, our \hyperref[sec:clean-test]{clean test protocol} uses verified dynamic positives as true labels and samples negatives only from the absolute-negative region outside the static candidate set.

To compare supervision sources fairly, we first restrict attention to the subset of binaries for which both static and dynamic labels are available. We split this common-source subset using the same package-level multi-label stratified procedure described above (approximately 80\% train, 10\% validation, and 10\% test), and then train two architecturally identical models: one under dynamic supervision and one under static supervision. Both are evaluated on the same clean test set, so any performance difference can be attributed to the supervision source rather than to architectural or evaluation differences. We then enlarge the static training set to the full static corpus while keeping the test set unchanged, which isolates the effect of training scale under the same evaluation protocol. This yields the three supervision regimes used later in Section~\ref{sec:RQ1}: \textit{dynamic}, \textit{static}, and \textit{large-scale static}.

\vspace*{2pt}
\noindent\textbf{\textit{Long-range test subset.}} 
To evaluate whether Dual Hubs help precisely where local message passing is weakest, we derive a \emph{long-range} subset from the test set. For each labeled pair, we compute the shortest-path distance between its source and destination on the \emph{base} ACFG, i.e., before adding any hub edges~\cite{NeuCall}. We classify a pair as \emph{long-range} if either: (1) its shortest path is greater than 4 hops, or (2) the source and destination are disconnected even though both endpoints are present in the recovered graph.

We define pairs separated by more than 4 hops, together with disconnected pairs, as long-range cases. This threshold is motivated by prior observations that the ability of message-passing GNNs to propagate and preserve task-relevant information degrades as graph distance increases~\cite{alon2021on,dwivedi2022LRGB}. Disconnected pairs represent the limiting case, as no information can be exchanged through local message passing. We therefore report performance on both the full test set (\emph{Overall}) and this long-range subset. The latter provides a targeted stress test of long-distance reasoning and allows us to assess whether the Dual Hubs improve information exchange beyond conventional local propagation.

\vspace*{2pt}
\noindent \textbf{\textit{Function-level deduplication.}} 
Existing datasets like Callee~\cite{Callee} and CupidCall~\cite{NeuCall} exhibit significant function-level overlap between training and test splits (\textbf{15.9\%} and \textbf{13.6\%}, respectively). This redundancy allows models to trivially memorize control-flow patterns, artificially inflating evaluation metrics. To rigorously evaluate true generalization, we enforce strict function-level deduplication. For each labeled function, we compute a SHA-256 hash over its sequence of assembly mnemonics to generate a structural fingerprint. After splitting, we strictly purge any test-set labels whose function fingerprint appears in the training set. This per-edge-type filtering removes 1.18\% to 15.32\% of overlapping test instances (Table~\ref{tab:gt-func-overlap-static-dup}), eliminating structural data leakage within the GAT's message-passing scope (details in Appendix~\ref{appendix:function-dedup}).

\definecolor{dyblue}{HTML}{1f77b4}
\definecolor{stgreen}{HTML}{2ca02c}
\definecolor{storange}{HTML}{ff7f0e}

\begin{figure*}[t]
    \centering
    \includegraphics[width=0.97\textwidth]{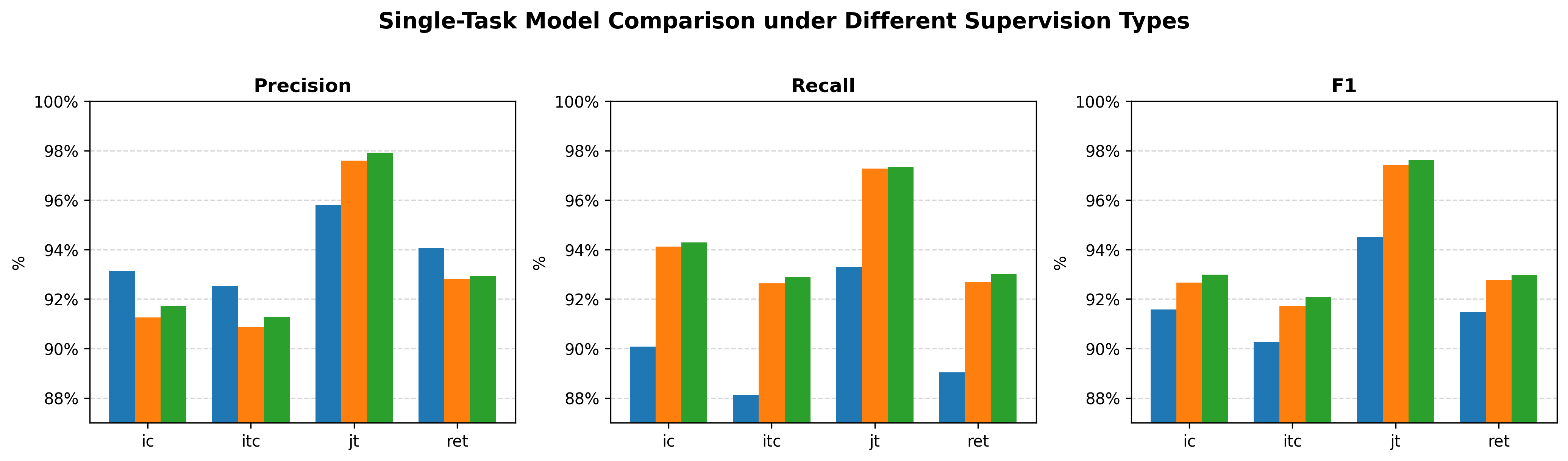}
    \vspace{-2mm}
    \caption{Performance comparison of single-task models utilizing Dual Virtual Hubs under dynamic
    \raisebox{-0.2ex}{\textcolor{dyblue}{\rule{0.9em}{0.9em}}},
    static \raisebox{-0.2ex}{\textcolor{storange}{\rule{0.9em}{0.9em}}},
    and large-scale static supervision across ICF types \raisebox{-0.2ex}{\textcolor{stgreen}{\rule{0.9em}{0.9em}}}. The x-axis corresponds, in order, to indirect calls, indirect tail calls, jump tables, and returns. Detailed results are provided in Appendix Table~\ref{tab:singlecomparison}.}
    \label{fig:single-model-comparison}
\vspace{-3mm}
\end{figure*}

\vspace*{2pt}
\noindent \textbf{\textit{Hyperparameter tuning.}} 
We utilize Optuna~\cite{optuna} to maximize validation F1 and identify a near-optimal configuration for our architecture. To ensure stable performance and fair comparability across all settings, we consistently apply the resulting configuration: 256 hidden units, 4 GAT layers, 8 attention heads, a dropout rate of 0.4, weight decay of $10^{-5}$, and a learning rate of 0.001. Comprehensive details regarding the hyperparameter search space and training regimen are deferred to Appendix~\ref{appendix:hyperparameters}.

\subsection{RQ1: Single-task performance with Dual Hubs}
\label{sec:RQ1}
\noindent To examine (1) the effect of dynamic versus static supervision on the same set of binaries and (2) the influence of training-set size, we evaluate single-task model performance incorporating the Dual Hub architecture under three supervision regimes (Figure~\ref{fig:single-model-comparison}):
\begin{enumerate}[leftmargin=12pt]
    \item Dynamic supervision (\raisebox{-0.2ex}{\textcolor{dyblue}{\rule{0.9em}{0.9em}}} bars): models trained on binaries whose ICF edges are derived from dynamic tracing;
    \item Static supervision (\raisebox{-0.2ex}{\textcolor{storange}{\rule{0.9em}{0.9em}}} bars): models trained on the same set of binaries but using compiler-level analysis labels in place of dynamic traces;
    \item Large-scale static supervision (\raisebox{-0.2ex}{\textcolor{stgreen}{\rule{0.9em}{0.9em}}} bars): models trained on the full static dataset, expanding the size of available training data.
\end{enumerate} 
All models are tested on a common clean test set, constructed from the test split of the dynamic dataset using our clean test protocol.  
The comparison between (1) dynamic and (2) static supervision isolates the effect of complementary ground truths, whereas the comparison between (2) static and (3) large-scale static supervision highlights how training-set size influences model performance. Note that in RQ1, we use Dual Hubs for all ICF types simply to isolate the effect of supervision sources. We evaluate and select the best architecture for each type (e.g., using no hubs for jump tables) later in RQ2.

\vspace*{2pt}
\noindent\textbf{\textit{Intra-procedural edges.}} Dynamic supervision performs significantly worse on jump-table edges because runtime traces seldom capture all dispatch targets, resulting in incomplete labels and lower overall precision, recall, and F1.

\vspace*{2pt}
\noindent\textbf{\textit{Inter-procedural edges.}} 
\noindent Figure~\ref{fig:single-model-comparison} compares static and dynamic supervision across three inter-procedural edge types. Trained on the same set of binaries, the two models exhibit complementary error profiles driven by their label sources. Dynamic supervision achieves higher precision because runtime traces include only verified execution edges, yielding very few false positives ($FP$). However, as many legitimate edges are never executed during tracing, the number of false negatives ($FN$) increases, leading to lower recall. In contrast, static supervision produces a broader set of positive candidates ($TP + FP$) by including all statically inferable control-flow edges. This broader coverage introduces more $FP$ during training but also captures many true positives ($TP$) that dynamic tracing misses, resulting in higher recall and F1. 
As shown in Figure~\ref{fig:single-model-comparison}, scaling the training data by 10$\times$ (\texttt{\_st\_large}) yields only marginal performance gains. This plateau suggests that the main bottleneck is label quality rather than data volume: simply adding more statically supervised data does not overcome the noise and bias inherent in static labels.

\begin{table}[t]
\centering
\caption{Ablation of the Dual Hub architecture on the long-range subset. All models are trained with the large-scale static supervision selected in RQ1.}
\vspace{-2mm}
\label{tab:dual-hub-long-range}
\resizebox{0.9\linewidth}{!}{
\begin{tabular}{lcccc}
\toprule
\textbf{Setting} & \textbf{Precision (\%)} & \textbf{Recall (\%)} & \textbf{F1 (\%)} & \textbf{\boldmath$\Delta F1$(\%)} \\
\midrule
\multicolumn{5}{c}{\textit{Indirect Call (ic)}} \\
Baseline   & 84.73 & 85.89 & 85.31 & -- \\
GCH        & 88.42 & 90.87 & 89.63 & \textbf{+4.32} \\
GCH + GDH  & 90.28 & 92.44 & 91.35 & \textbf{+6.04} \\
\addlinespace[-0.5mm]
\multicolumn{5}{c}{\textit{Indirect Tail Call (itc)}} \\
Baseline   & 80.61 & 81.85 & 81.23 & -- \\
GCH        & 86.72 & 88.35 & 87.53 & \textbf{+6.30} \\
GCH + GDH  & 89.54 & 91.19 & 90.36 & \textbf{+9.13} \\
\addlinespace[-0.5mm]
\multicolumn{5}{c}{\textit{Jump Table (jt)}} \\
Baseline   & 97.35 & 97.68 & 97.51 & -- \\
GCH        & 96.12 & 96.45 & 96.28 & \textbf{-1.23} \\
GCH + GDH  & 95.10 & 95.64 & 95.37 & \textbf{-2.14} \\
\addlinespace[-0.5mm]
\multicolumn{5}{c}{\textit{Return (ret)}} \\
Baseline   & 86.27 & 86.41 & 86.34 & -- \\
GCH        & 91.13 & 90.75 & 90.94 & \textbf{+4.60} \\
GCH + GDH  & 91.60 & 91.16 & 91.38 & \textbf{+5.04} \\
\bottomrule
\end{tabular}%
}
\vspace{-5mm}
\end{table}

\subsection{RQ2: Long Range ICF Pairs Analysis}
\label{sec:  Long distance ICF pairs analysis}

As shown in Table~\ref{tab:dual-hub-long-range}, RQ2 isolates the structural contribution of the Dual Virtual Hubs on the long-range subset defined in Section~\ref{sec:setup}. We compare three variants: \texttt{Baseline} (no hubs), \texttt{GCH} (code-side hub only), and \texttt{GCH + GDH} (the full Dual Hub architecture).

\vspace*{2pt}
\noindent\textbf{\textit{Inter-procedural tasks.}}
For the three inter-procedural tasks---indirect calls, indirect tail calls, and returns---the baseline is clearly limited by local propagation, with F1 ranging from 81.23\% to 86.34\%. Adding the Global Code Hub already yields substantial gains (\textit{ic}: +4.32, \textit{itc}: +6.30, \textit{ret}: +4.60), indicating that code-side long-range routing is useful when the source and target are far apart or disconnected in the base ACFG. Adding the Global Data Hub brings a further improvement for all three tasks, raising the total gains to +6.04 for \textit{ic}, +9.13 for \textit{itc}, and +5.04 for \textit{ret}. The largest improvement appears on \textit{itc}, suggesting that this task benefits most from combining code-side reachability with additional data-side evidence.

\vspace*{2pt}
\noindent\textbf{\textit{Jump tables.}}
The behavior is different for jump tables. Because jump-table resolution is predominantly intra-procedural and depends on localized arithmetic patterns (e.g., base-plus-offset style dispatch), the baseline already performs strongly at 97.51\% F1. Adding global routing slightly hurts performance (-1.23 with GCH and -2.14 with GCH + GDH), suggesting that hub-based aggregation introduces unnecessary inter-procedural noise for this task.

\vspace*{2pt}
\noindent\textbf{\textit{Takeaway.}}
These results support a task-aware structural design. Dual Hubs are beneficial for inter-procedural tasks (\textit{ic}, \textit{itc}, and \textit{ret}), where long-range routing is often necessary, whereas jump-table prediction is better kept local. We therefore carry this per-type architectural choice forward in RQ3: inter-procedural tasks retain hub routing, while jump-table targets remain outside global routing.

\subsection{RQ3: Multi-Task Model (MTL) Performance}
\label{sec: MTL Performance}


As Figure~\ref{fig:single-model-comparison} shows, scaling the static training set by 10$\times$ yields only a marginal 0.27\% average F1 improvement. This saturation suggests that single-task models have largely exhausted the gains obtainable from simply adding more static supervision. Motivated by RQ2, we therefore introduce multi-task learning with \emph{task-aware routing}: inter-procedural tasks (\textit{ic}, \textit{itc}, \textit{ret}) retain hub-based routing, while jump-table targets remain local. This design encourages cross-type transfer without forcing highly localized tasks to absorb unnecessary global context. Tables~\ref{tab:mtl-performance} and~\ref{tab:mtl-overall} show that MTL consistently improves over the corresponding single-task baselines.

\begin{table}[t]
\centering
\caption{Performance comparison of Single-Task, Multi-Task (MTL), and the auxiliary type-prediction variant (MTL\_t) on the \textit{Long-range test dataset}. All models are trained on the large-scale static dataset. \textit{Note:} \textsuperscript{\dag}Dual hub setting; \textsuperscript{\ddag}No hub setting.}
\vspace{-2mm}
\label{tab:mtl-performance}
\resizebox{0.95\linewidth}{!}{
\begin{tabular}{lccccc}
\toprule
\textbf{Task Type} & \textbf{Setting} & \textbf{Precision (\%)} & \textbf{Recall (\%)} & \textbf{F1 (\%)} & \textbf{\boldmath$\Delta F1$(\%)} \\
\midrule
\multirow{3}{*}{\textit{Indirect Call (ic)}} 
& Single\textsuperscript{\dag}  & 90.28 & 92.44 & 91.35 & -- \\
& MTL                           & 94.68 & \textbf{96.35} & 95.51 & +4.16 \\
& MTL\_t              & \textbf{95.36} & 96.30 & \textbf{95.83} & \textbf{+4.48} \\
\midrule
\multirow{3}{*}{\textit{Indirect Tail Call (itc)}} 
& Single\textsuperscript{\dag}  & 89.54 & 91.19 & 90.36 & -- \\
& MTL                           & \textbf{96.15} & \textbf{96.19} & \textbf{96.17} & \textbf{+5.81} \\
& MTL\_t               & 95.81 & 95.17 & 95.49 & +5.13 \\
\midrule
\multirow{3}{*}{\textit{Jump Table (jt)}} 
& Single\textsuperscript{\ddag} & 97.35 & 97.68 & 97.51 & -- \\
& MTL                           & 97.05 & \textbf{98.01} & 97.53 & +0.02 \\
& MTL\_t               & \textbf{98.25} & 97.65 & \textbf{97.95} & \textbf{+0.44} \\
\midrule
\multirow{3}{*}{\textit{Return (ret)}} 
& Single\textsuperscript{\dag}  & 91.60 & 91.16 & 91.38 & -- \\
& MTL                           & \textbf{96.81} & \textbf{96.81} & \textbf{96.81} & \textbf{+5.43} \\
& MTL\_t               & 96.11 & 96.47 & 96.29 & +4.91 \\
\bottomrule
\end{tabular}%
}
\vspace{-3mm}
\end{table}

\begin{table}
\centering
\caption{Performance comparison of Single-Task, Multi-Task (MTL), and MTL with type prediction (MTL\_t) on the \textit{Overall test dataset}. Same training setting and notation as Table~\ref{tab:mtl-performance}.}
\vspace{-2mm}
\label{tab:mtl-overall}
\resizebox{0.95\linewidth}{!}{
\begin{tabular}{lccccc}
\toprule
\textbf{Task Type} & \textbf{Setting} & \textbf{Precision (\%)} & \textbf{Recall (\%)} & \textbf{F1 (\%)} & \textbf{\boldmath$\Delta F1$(\%)} \\
\midrule
\multirow{3}{*}{\textit{Indirect Call (ic)}} 
& Single\textsuperscript{\dag}  & 91.73 & 94.28 & 92.99 & -- \\
& MTL                           & \textbf{96.55} & 95.82 & 96.18 & +3.19 \\
& MTL\_t               & 95.82 & \textbf{96.85} & \textbf{96.33} & \textbf{+3.34} \\
\midrule
\multirow{3}{*}{\textit{Indirect Tail Call (itc)}} 
& Single\textsuperscript{\dag}  & 91.29 & 92.88 & 92.08 & -- \\
& MTL                           & 97.28 & \textbf{96.32} & \textbf{96.80} & \textbf{+4.72} \\
& MTL\_t               & \textbf{97.66} & 95.18 & 96.40 & +4.32 \\
\midrule
\multirow{3}{*}{\textit{Jump Table (jt)}} 
& Single\textsuperscript{\ddag} & 98.63 & \textbf{98.21} & 98.42 & -- \\
& MTL                           & \textbf{98.95} & 98.01 & \textbf{98.48} & \textbf{+0.06} \\
& MTL\_t               & 98.65 & 97.97 & 98.31 & -0.11 \\
\midrule
\multirow{3}{*}{\textit{Return (ret)}} 
& Single\textsuperscript{\dag}  & 92.93 & 93.01 & 92.97 & -- \\
& MTL                           & 98.05 & \textbf{96.55} & \textbf{97.29} & \textbf{+4.32} \\
& MTL\_t               & \textbf{98.12} & 95.80 & 96.95 & +3.98 \\
\bottomrule
\end{tabular}%
}
\vspace{-5mm}
\end{table}

\noindent \textbf{\textit{Efficacy of MTL.}} 
On the \textit{Long-range} subset (Table~\ref{tab:mtl-performance}), where local message passing is most limited, the gains are largest for the three inter-procedural tasks: +4.16 F1 for \textit{ic}, +5.81 for \textit{itc}, and +5.43 for \textit{ret}. These improvements indicate that, once Dual Hubs provide the necessary structural reachability, shared training can further transfer useful semantics across related ICF types. For jump tables, the gains are much smaller because this task is already highly localized and near saturation; nevertheless, MTL does not degrade performance, and MTL\_t yields a small improvement (+0.44 F1) on the long-range subset.
On the \textit{Overall} test set (Table~\ref{tab:mtl-overall}), the same trend remains, although the relative gains are smaller because the distribution is dominated by easier local pairs. MTL still improves F1 over the single-task baseline for all four ICF types. For jump tables, all three variants operate in a near-ceiling regime (98.31\%--98.48\% F1). Here, MTL remains slightly better than the single-task baseline, while MTL\_t shows a minor trade-off between auxiliary type prediction and task-specific optimization. We therefore treat \textbf{MTL} as the most stable overall predictive model, and \textbf{MTL\_t} as an  auxiliary variant that encourages subtype-discriminative representations rather than uniformly maximizing every per-task metric.

\begin{table}[t]
\centering
\caption{Performance on the Callee~\cite{Callee}, SJA~\cite{nguyen2024}, and our ICFlowNet datasets. \textit{Note: MTL is evaluated only on our dataset because Callee and SJA each cover a single ICF type.}}
\vspace{-2mm}
\label{tab:callee-sja-results}
\resizebox{0.48\textwidth}{!}{%
\begin{tabular}{llccc}
\toprule
\textbf{Dataset} & \textbf{Model} & \textbf{Precision (\%)} & \textbf{Recall (\%)} & \textbf{F1 (\%) (\boldmath{$\Delta$}\textbf{F1})} \\
\midrule
\multirow{3}{*}{\textbf{Callee}} 
 & Callee         & 91.85 & 90.00 & 90.92 (\textbf{$-6.60$}) \\
 & CupidCall      & 94.26 & 92.57 & 93.41 (\textbf{$-4.11$}) \\
 & \textbf{Our STL} & 98.22 & 96.83 & \textbf{97.52} (---) \\
\midrule
\multirow{3}{*}{\shortstack[l]{\textbf{ICFlowNet}\\ \textbf{(Overall)}}} 
 & Callee         & 88.12 & 88.62 & 88.37 (\textbf{$-7.81$}) \\
 & CupidCall      & 91.03 & 89.35 & 90.18 (\textbf{$-6.00$}) \\
 & \textbf{Our MTL}  & 96.55 & 95.82 & \textbf{96.18} (---) \\
\midrule
\multirow{3}{*}{\shortstack[l]{\textbf{ICFlowNet}\\ \textbf{(Long-range)}}} 
 & Callee         & 80.25 & 84.37 & 82.26 (\textbf{$-13.25$}) \\
 & CupidCall      & 81.16 & 83.84 & 82.48 (\textbf{$-13.03$}) \\
 & \textbf{Our MTL}  & 94.68 & 96.35 & \textbf{95.51} (---) \\
\midrule
\multirow{2}{*}{\textbf{SJA}}    
 & SJA            & 97.43 & 99.80 & 98.60 (\textbf{$-0.89$}) \\
 & \textbf{Our STL} & 99.28 & 99.70 & \textbf{99.49} (---) \\
\midrule
\multirow{2}{*}{\shortstack[l]{\textbf{ICFlowNet}\\ \textbf{(Overall)}}}    
 & SJA            & 95.35 & 94.32 & 94.83 (\textbf{$-3.65$}) \\
 & \textbf{Our MTL} & 98.95 & 98.01 & \textbf{98.48} (---) \\
\midrule
\multirow{2}{*}{\shortstack[l]{\textbf{ICFlowNet}\\ \textbf{(Long-range)}}}    
 & SJA            & 94.56 & 93.17 & 93.86 (\textbf{$-3.67$}) \\
 & \textbf{Our MTL} & 97.05 & 98.01 & \textbf{97.53} (---) \\
\bottomrule
\end{tabular}%
}
\vspace{-5mm}
\end{table}


As a control, Appendix~\ref{appendix:MTL Performance withoutt Dual Hubs} shows that removing the Dual Hubs reduces the average MTL gain to only 1.33\%, indicating that task sharing alone is insufficient. We also refer the reader to Appendix~\ref{appendix:Feature level fusion}, where a simple feature-fusion baseline shows some benefit on data-rich tasks, but unlike our architecture, it does not consistently improve all task types. Finally, Appendix~\ref{appendix:Optimization_Robustness} shows that the MTL model remains robust across compiler optimization levels O0--O3.

\vspace*{2pt}
\noindent \textbf{\textit{Recall improvements.}} 
Beyond F1, MTL also improves recall for the inter-procedural tasks, especially on the long-range subset. This is important for binary analysis, where missing valid targets is often more harmful than introducing a small number of extra candidates. Intuitively, shared training allows abundant tasks such as \textit{ret} to regularize rarer tasks such as \textit{itc}, reducing false negatives while preserving strong precision.

\subsection{RQ4: Comparative Evaluation}
\label{sec: Comparative Study} 


\noindent \textbf{\textit{Comparison scope.}}
Since prior studies do not cover \textit{ret} and \textit{tailcall} edges, we restrict our direct baseline comparison in the main text to indirect calls and jump tables. For indirect calls, we focus on \emph{Callee}~\cite{Callee} and \emph{CupidCall}~\cite{NeuCall}, which are the closest learned baselines in task formulation and available reproducibility. We 
do not include \emph{AttnCall}~\cite{AttnCall2024} in the main empirical baseline set because its released artifacts do not support a controlled reproduction, and
its reported setup is not aligned with real binary indirect-call supervision
~\cite{NeuCall}. 
We do not treat \emph{TypeArmor}~\cite{TypeArmor}, \emph{TypeSqueezer}~\cite{Lin2023TypeSqueezer}, \emph{BPA}~\cite{BPA}, \emph{SchedExec}~\cite{Shi2024}, \emph{BinDSA}~\cite{Gao2025}, or \emph{Disa}~\cite{Wang2025} as direct baselines, because they make materially different assumptions and produce different outputs, ranging from signature-based pruning and pointer-analysis-based call-graph refinement to disassembly-assisted CFG recovery. Among them, 
we reserve the closest recent program-analysis systems~\cite{Gao2025,Wang2025} for external comparison in Appendix~\ref{appendix:External_Comparison} rather than claim a strict artifact-level apples-to-apples reproduction. For jump tables, we compare directly against the state-of-the-art analyzer \emph{SJA}~\cite{nguyen2024}.


\vspace*{2pt}
\noindent \textbf{\textit{Evaluation methodology.}} We evaluate prior baselines in two settings. First, on the original Callee~\cite{Callee} and SJA~\cite{nguyen2024} datasets, we assess our single-task model (STL) under the same benchmark conventions, while enforcing strict function-level deduplication. Second, on our dataset, we compare our multi-task model against these baselines on both the \textit{Overall} test set and the structurally harder \textit{Long-range} subset. Table~\ref{tab:callee-sja-results} reports the results.


\vspace*{2pt}
\noindent \textbf{\textit{Indirect call prediction.}} 
On the original Callee dataset, strict deduplication lowers the reported performance of prior baselines: Callee and CupidCall reach 90.92\% and 93.41\% F1, respectively, whereas our STL attains 97.52\%. On our \textit{Overall} dataset, the gap widens further: our MTL reaches 96.18\% F1, compared with 88.37\% for Callee and 90.18\% for CupidCall. The difference is largest on the \textit{Long-range} subset, where Callee and CupidCall fall to 82.26\% and 82.48\% F1, while our model maintains 95.51\%. These results suggest that direct-call-transfer and local graph propagation alone are insufficient for difficult long-range cases, whereas our Dual-Hub design remains effective when the source and target are far apart or disconnected in the base graph.


\vspace*{2pt}
\noindent \textbf{\textit{Jump table prediction.}} 
On SJA's original dataset, our STL slightly exceeds SJA (99.49\% vs.\ 98.60\% F1). On our dataset, however, the gap becomes larger: SJA reaches 94.83\% F1 on the \textit{Overall} set and 93.86\% on the \textit{Long-range} subset, whereas our MTL achieves 98.48\% and 97.53\%, respectively. A likely reason is compiler drift: SJA was developed and evaluated on older lowering patterns, while our LLVM 19 binaries contain more aggressive transformations, including merged switch constructs, that are harder to capture with fixed rules. In contrast, our learned model adapts better to these newer structural patterns.

\subsection{RQ5: Efficiency and Scalability}
\label{sec:rq5-efficiency}

Beyond predictive accuracy, we evaluate the practical cost of the Dual Virtual Hubs and the multi-task learning (MTL) design.

\begin{table}[t]
\centering
\caption{Topological overhead introduced by the Dual Virtual Hubs, measured as edge inflation relative to the pruned base ACFG.}
\vspace{-2mm}
\label{tab:hub_topology}
\resizebox{0.45\textwidth}{!}{%
\begin{tabular}{lccc}
\toprule
\textbf{Task Type} & \textbf{GCH (\%)} & \textbf{GDH (\%)} & \textbf{Total (\%)} \\
\midrule
Indir. Call & +1.70 & +3.28 & +4.98 \\
Ret & +2.13 & +3.96 & +6.09 \\
Indir. Tail Call$^\dagger$ & +1.92 & +3.73 & +5.65 \\
\midrule
MTL (Unified)$^\ddagger$ & +3.72 & +7.72 & +11.44 \\
\bottomrule
\multicolumn{4}{p{0.95\linewidth}}{\small \textit{Note:} $^\dagger$ Tail-call candidates include ambiguous indirect jumps unresolved by \textit{angr}. \newline $^\ddagger$ MTL is measured on binaries containing at least one inter-procedural edge.}
\end{tabular}%
}
\vspace{-4mm}
\end{table}

\vspace*{2pt}
\noindent \textbf{\textit{Topological overhead of Dual Virtual Hubs.}} 
A naive dense routing scheme that connects the GCH to all code nodes and the GDH to all data nodes would substantially enlarge the graph. In contrast, our candidate-aware routing keeps the added edges modest: +4.98\% for indirect calls, +6.09\% for returns, and +5.65\% for indirect tail calls (Table~\ref{tab:hub_topology}). Importantly, this overhead does not stack linearly in the unified MTL graph. Because the inter-procedural tasks share many code candidates and nearby data nodes, the same hubs can be reused across tasks, so the total overhead remains only +11.44\%. These results indicate that Dual Hubs provide long-range routing while keeping graph growth limited.

\vspace*{2pt}
\noindent \textbf{\textit{Temporal efficiency of MTL.}} 
We compare one unified MTL model against a suite of four independent single-task (ST) models. By sharing the lower GAT layers and executing a single data pipeline per graph, MTL reduces cumulative per-graph training time from 9.19\,s to 4.08\,s ($\approx 2.25\times$ speedup). The benefit is even clearer at inference time: the ST baseline requires four separate forward passes, whereas MTL requires only one, reducing per-graph latency from 3.79\,s to 1.57\,s. Thus, the multi-task design improves not only predictive performance but also end-to-end throughput.

\section{Discussion and Conclusion}
\label{sec:discussion}

\noindent \textbf{\textit{Design trade-offs and remaining scalability challenges.}}
Our dual-hub design provides a compact compromise between long-range routing and structural sparsity. The GCH shortens distant code-side dependencies, while the GDH aggregates nearby data evidence that may be missing from direct xRef chains. This design is intentionally minimal: adding more hubs could distribute information further, but might also lengthen routing paths and weaken the distance-compression effect that makes virtual routing useful in the first place. More broadly, our results suggest that long-range binary graph reasoning is governed by a tension between two competing forces: shortening communication distance and avoiding excessive information mixing. Pushing too far toward global routing risks introducing structural noise, especially for highly localized mechanisms such as jump tables; pushing too far toward local sparsity, however, leaves distant inter-procedural evidence unreachable. In addition, although candidate-aware routing already keeps graph growth modest, memory remains a practical concern for very large binaries. A natural next step is candidate-centric graph compression, where the model encodes only task-relevant subgraphs around candidate source/destination sets and their data neighborhoods.

\vspace*{2pt}
\noindent \textbf{\textit{Limitations for security-critical deployment.}}
ICFlowNet is designed as a high-fidelity prediction framework, but it is not a universal replacement for conservative program-analysis pipelines. Like other neural approaches, it depends on the training distribution and inherits some optimization variability across datasets and runs. Although our evaluation shows strong and stable performance, these properties can limit direct applicability to safety-critical tasks that require strict completeness guarantees. A representative example is fine-grained control-flow integrity (CFI), where even a small number of missed legitimate targets may be unacceptable. In such settings, ICFlowNet is better viewed as a complementary component---for example, to prioritize candidates, improve binary understanding, shrink search spaces, or augment conservative analyses---rather than as a drop-in substitute for methods that require a hard 100\% recall guarantee.

\vspace*{2pt}
\noindent \textbf{\textit{Conclusion.}}
ICFlowNet is a unified framework for indirect control-flow prediction in stripped binaries. By combining Dual Virtual Hubs, multi-task heterogeneous graph learning, and a leakage-aware evaluation pipeline, it improves both robustness and generalization. Our experiments show that structural repair and cross-type transfer are both important for accurate long-range ICF recovery.


\bibliographystyle{unsrt}
\bibliography{bib/ibranch, bib/path-exploration,bib/binary-code,bib/code-reuse,bib/debloating,bib/mips,bib/other,bib/jiang,bib/others,bib/gnn,bib/binary,bib/asplos}

\appendix 

\section{Open Science}
The artifact includes code for graph construction, training, and evaluation, with a separate package containing binaries, metadata, and model resources. The code is available on \href{https://github.com/Ryan-hub-bit/icflownet_artifact}{GitHub}.

\section{Ethical Considerations}

This work studies indirect control-flow prediction for binary analysis. Our goal is to improve the completeness and fidelity of recovered control-flow graphs for defensive and scientific uses, including program understanding, binary hardening, and evaluation of downstream analysis tools. At the same time, we recognize that binary-analysis techniques are inherently dual-use: more accurate recovery of indirect control-flow edges could also be used to support offensive reverse engineering or to improve the precision of attack planning against existing binaries.

We mitigate this risk in several ways. First, our paper does not introduce a new exploit, attack payload, bypass, or vulnerability trigger, nor does it disclose previously unknown vulnerabilities in deployed software. The contribution is a prediction and evaluation framework for control-flow recovery. Second, all binaries in our dataset are built from publicly available open-source software, and our experiments are conducted in an offline research setting on locally compiled binaries. We do not instrument or probe third-party production systems, and we do not collect private user data, credentials, or telemetry from end users. Third, our dynamic tracing is used only to collect execution traces necessary for ground-truth construction and evaluation; it is not used to tamper with software behavior in deployed environments. Overall, we believe the defensive and scientific benefits outweigh the limited dual-use risk.




\section*{Acknowledgments}

We thank all CCS anonymous reviewers for their valuable comments to improve this paper. 
This work is supported by NSF grants 2417055 \& 2555093 and Google Research Scholar Award.


\setcounter{table}{0}
\setcounter{figure}{0}
\renewcommand{\thetable}{A\arabic{table}}
\renewcommand{\thefigure}{A\arabic{figure}}

\section*{Appendix}

\section{Indirect Control Flow Categorization} 
\label{sec:appendix-categorization}

In deep learning-based approaches to indirect control-flow prediction, the existence of an edge between a source and destination basic block is inferred from their instruction semantics—typically captured via learned embeddings. This differs from traditional rule-based methods that rely on symbolic matching or static patterns. Consequently, the way we categorize edge types must align with patterns that are learnable from instruction-level representations.

\vspace{2pt}
\noindent
\textbf{\textit{Three indirect jump subtypes.}}
In our formulation, we divide the general category of indirect jumps into three subtypes—jump table, indirect tail call, and return---each treated as a distinct prediction task. Although all three manifest as indirect jump-based instructions at the binary level, their semantic origins, structural patterns, and frequencies differ significantly. Jump tables are compiler-generated constructs for implementing switch-case statements, resulting in intra-procedural, data-dependent control flow through a dispatch table. Indirect tail calls emerge from tail-call optimization, where a function concludes by jumping to another without allocating a new stack frame, yielding inter-procedural transfers that mimic calls without returns. Return edges restore control after function execution, typically executed as \texttt{pop rax; jmp rax}, and depend on stack-based historical context to resume execution after a prior call.

These subtypes also differ in their structural characteristics: jump table targets generally reside within the same function, indirect tail calls jump across function boundaries, and returns reconnect to the caller’s successor block. These distinctions give rise to unique instruction embeddings and control-flow signatures. Moreover, their frequency distributions are highly imbalanced---returns appear far more often than jump tables, and jump tables more than indirect tail calls. Modeling them as a single category would bias the model toward the majority class, reducing predictive accuracy on rarer yet semantically critical constructs like indirect tail calls. To account for their distinct semantic roles, structural behaviors, and distributional properties, we treat these indirect jump subtypes as separate edge types.

\vspace{2pt}
\noindent
\textbf{\textit{Unified treatment of indirect calls.}}  
In traditional static analysis and control-flow integrity techniques, \texttt{indirect calls} are often divided into three categories: function pointer dispatch, virtual table dispatch, and dynamic method lookup, based on how the call target is computed. This classification helps enforce fine-grained control policies.

In contrast, in our learning-based approach, all indirect calls exhibit highly consistent patterns at the assembly level: the source is a \texttt{call} instruction to a register or memory operand (with varying addressing patterns), and the destination is always a function entry point. These structural similarities lead to strong overlap in instruction-level features for both source and destination. More importantly, since our heterogeneous graph includes data nodes and cross-reference edges between data nodes and code nodes, the model can reason about different dispatch patterns contextually. We therefore treat all indirect calls as a unified edge type, which simplifies supervision while preserving the ability to learn nuanced call resolution behavior.



\section{Static ICF Pairs Collection} \label{sec:appendix-static-icf}

We extend Clang to insert callee function type annotations at each indirect callsite,  based on LLVM-CFI~\cite{LLVM-CFI}. These type annotations are essential for resolving indirect calls and indirect tail calls, as they enable precise identification of valid callee functions. A custom LLVM pass is then integrated into the x86 backend to encode the dispatch and target positions of all ICF pairs using our instruction labeling system, ensuring accurate extraction by the final parser.

For switch-case jump tables, we leverage LLVM’s \texttt{X86 DAG{\textrightarrow}DAG Instruction Selection} pass, which explicitly emits jump table constructs and records both dispatch sites and valid target addresses. For indirect calls and indirect tail calls, we adopt a type-based strategy inspired by Modular CFI~\cite{NiuTan2014MCFI}. Specifically, callee type annotations are propagated from Clang through the Machine IR (MIR) stage, where potential callees are tagged with their function signatures. The candidate set is then restricted according to LLVM’s internal criteria—namely, functions that are externally visible or have their addresses taken. Customized parsing scripts subsequently match each indirect callsite to its valid callees using these propagated type signatures.

Finally, return edges are reconstructed based on the history of indirect calls. For returns originating from indirect calls, we trace the call stack to associate each return site with the basic block immediately following its corresponding callsite. In the case of indirect tail calls---where the callee does not return to the immediate caller---we identify the original caller of the function containing the tail call and link the return to the first basic block following the callsite in that original caller. This approach enables accurate recovery of return edges, even in the presence of tail-call optimization.

\section{Detailed Dataset Construction Pipeline}\label{appendix:dataset}

We begin by compiling a large corpus of binaries from the Arch Linux ecosystem (version 2024.02.01), encompassing all packages from the official core and extra repositories, along with a curated subset from the Arch User Repository (AUR). In total, we selected 12,972 unique C/C++ projects, of which approximately 77\% were successfully compiled. Each package was built from source using a custom LLVM-based toolchain (version 19.1.5) and integrated with Arch Linux’s native makepkg and PKGBUILD infrastructure.

Following compilation, we deduplicate binaries using a normalized binary hash to eliminate redundant outputs. We retain only those binaries that contain at least one static ICF pair, extracting them using a custom instruction-labeling parser. To establish a dynamic ground truth, we additionally apply a dynamic tracing pipeline for packages whose PKGBUILD includes a check() function (i.e., native test suites), which successfully annotated 1,351 binaries.

\vspace{2pt}
\noindent
\textbf{\textit{LLM-driven dynamic test augmentation.}} Although native test suites provide an essential foundation, we identified a critical coverage bias when relying exclusively on them. For medium-to-large applications commonly encountered in everyday computing, native tests often fail to adequately exercise complex indirect control flow paths, resulting in a sparse yield of dynamic ICF pairs. To construct a fairer and more representative evaluation baseline, we deliberately selected 20 ubiquitous medium-sized real-world projects (e.g., PostgreSQL, Wireshark) from the initial pool of 1,351 binaries. 
For selected complex applications, we leveraged Large Language Models (LLMs) to automatically generate supplementary test suites. This engineering choice was motivated by recent work showing that LLM-based test generation can improve code coverage and steer execution toward previously uncovered branches or more complex execution paths~\cite{Ryan2024SymPrompt,Wang2024HITS,Jiang2024DirectedInput}. 
In our setting, these deeper executions 
increased the number of unique dynamically observed ICF pairs by approximately 25\% over native tests alone.
By enriching the dynamic ground truth for these heavy-duty applications, our dataset prevents subsequent evaluations from over-indexing on trivial paths found in smaller utilities, ensuring a benchmark that accurately reflects real-world software complexity.

\section{Package-Level Stratified Data Splitting} \label{appendix:data-split}

To obtain fair, representative, and leakage-resistant data splits for ICF prediction, we adapt multi-label proportional sampling~\cite{MultiLabel} to the package level. We treat each \emph{package} as the split unit, and assign to it a multi-label vector indicating whether any binary produced by that package contains at least one edge of each ICF type. This choice is important because binaries from the same package often share code templates, compiler options, and dependencies. Assigning the entire package---rather than individual binaries or graphs---to a single split therefore helps prevent cross-split information leakage.

Our splitting procedure is rarity-aware. We process edge types from rarest to most common so that low-frequency labels are not underrepresented in evaluation. For each edge type, we consider packages containing that label in randomly shuffled order and greedily assign each package to the split that is currently most under-filled for that label, targeting approximate per-type ratios of 80\% training, 10\% validation, and 10\% testing. We prioritize \textit{test}~$\rightarrow$~\textit{validation}~$\rightarrow$~\textit{train} during this process to ensure that the evaluation splits receive sufficient positive coverage even for scarce edge types.

Because a package may contain multiple binaries and multiple ICF types, each assignment may simultaneously affect several labels. The greedy procedure therefore respects both multi-label co-occurrence and package-level isolation. Once the evaluation targets are satisfied, the remaining packages are assigned to the training split. In this way, we preserve package-level independence while maintaining balanced per-type coverage across train, validation, and test sets.

Finally, we reuse the same package-level split for both single-task and multi-task experiments. This ensures that each single-task model is evaluated under conditions that are statistically consistent with the multi-task setting, enabling fair comparison while reducing the risk of type skew and cross-split leakage.

\section{Function-Level Deduplication for Leakage-Free Evaluation} 
\label{appendix:function-dedup}
To ensure a fair evaluation, we split the Callee~\cite{Callee} dataset according to the prescribed proportions while also enforcing a package-level split to reduce information leakage. Despite this effort, we observed that approximately \textbf{15.9\%} of ground truth-labeled functions overlapped between the training and testing sets.  And we further found that CupidCall~\cite{NeuCall} exhibited a similar issue with \textbf{13.6\%} overlap. This indicates that \textit{identical functions containing control-flow labels are frequently duplicated across splits}. Such redundancy poses a significant issue, as it enables the model to memorize function behaviors during training and trivially reproduce them during testing. Consequently, this memorization effect can artificially inflate performance metrics, undermining the validity of the evaluation.
\begin{table}[t]
\centering
\caption{Overlap rate of source ground truth-labeled functions between training and test splits (dynamic vs. static).}
\label{tab:gt-func-overlap-static-dup}
\begin{tabular}{l|r|r}
\hline
\textbf{Edge Type} & \textbf{Dynamic (\%)} & \textbf{Static (\%)} \\
\hline

Indir. Call      & 15.32 & 12.35 \\
Indir. Tail Call & 1.18  & 1.46  \\
Jump Table       & 8.23  & 9.25  \\
Return           & 12.36 & 14.75 \\
\hline
\end{tabular}
\vspace{-4mm}
\end{table}

To prevent structural data leakage in inductive evaluation, we apply function-level deduplication, excluding any ground truth edge whose surrounding control-flow context---the function in which it resides---shares a structural fingerprint with one in the training set. Specifically, for each \textit{ground-truth-labeled function} (i.e., a function containing at least one labeled source address) encountered during ACFG construction, we extract its instruction mnemonics from basic blocks, concatenate them in control-flow order, and compute a SHA-256 hash of the resulting string. This hash serves as a content-based structural fingerprint that remains consistent across duplicated code. The hashing is performed independently for each edge type, and the results are stored in per-binary JSON files (e.g., \texttt{\{binary\}\_retfunchashes.json}). By enforcing this constraint, we ensure that no test label is structurally equivalent to any training example within the GAT’s message-passing radius, thereby preserving the validity of generalization evaluation.

\section{Hyperparameter Tuning } 
\label{appendix:hyperparameters}

Optuna~\cite{optuna} is used to maximize validation F1 across all single-task models, and training is conducted for up to 50 epochs with early stopping (patience = 15), using the Adam optimizer~\cite{adam} and a fixed prediction threshold of 0.5. The hyperparameter search spans hidden dimensions ${64, 128, 256, 512}$, attention heads ${4, 8, 16}$, GAT layers ${2, 3, 4, 5}$, dropout rates ${0.1, 0.2, 0.3, 0.4, 0.5}$, weight decay ${10^{-6}, 10^{-5}, 10^{-4}, 10^{-3}}$, and learning rates ${10^{-4}, 10^{-3}, 10^{-2}}$. A near-optimized 
configuration---256 hidden units, 4 GAT layers, 8 attention heads, dropout of 0.4, weight decay of $10^{-5}$, and a learning rate of 0.001---is adopted for training and evaluation across all single-task models, demonstrating stable and generalizable performance on both static and dynamic datasets. This configuration is applied consistently in all experiments to ensure comparability and fair evaluation across training regimes.

\begin{table}[t]
\centering
\caption{Performance comparison of single-task models trained under dynamic, static, and large-scale static supervision across ICF types.}
\label{tab:singlecomparison}
\setlength{\tabcolsep}{6pt} 
\small 
\begin{tabular}{lccc}
\toprule
\textbf{Setting} & \textbf{Precision (\%)} & \textbf{Recall (\%)} & \textbf{F1 (\%)} \\
\midrule
\multicolumn{4}{c}{\textit{Indirect Call (ic)}} \\
ic\_dy        & \textbf{93.13} & 90.08 & 91.58 \\
ic\_st        & 91.26 & 94.12 & 92.67 \\
ic\_st\_large & 91.73 & \textbf{94.28} & \textbf{92.99} \\
\addlinespace[2mm]
\multicolumn{4}{c}{\textit{Indirect Tail Call (itc)}} \\
itc\_dy       & \textbf{92.53} & 88.12 & 90.27 \\
itc\_st       & 90.85 & 92.63 & 91.73 \\
itc\_st\_large & 91.29 & \textbf{92.88} & \textbf{92.08} \\
\addlinespace[2mm]
\multicolumn{4}{c}{\textit{Jump Table (jt)}} \\
jt\_dy        & 95.78 & 93.29 & 94.52 \\
jt\_st        & 97.59 & 97.27 & \textit{97.43} \\
jt\_st\_large & \textbf{97.92} & \textbf{97.34} & \textbf{97.63} \\
\addlinespace[2mm]
\multicolumn{4}{c}{\textit{Return (ret)}} \\
ret\_dy       & \textbf{94.07} & 89.03 & 91.48 \\
ret\_st       & 92.81 & 92.69 & 92.75 \\
ret\_st\_large & 92.93 & \textbf{93.01} & \textbf{92.97} \\
\bottomrule
\end{tabular}
\vspace{1mm}

\caption*{\textit{Note.} Each block compares models trained with dynamic (\texttt{\_dy}), static (\texttt{\_st}), and large-scale static (\texttt{\_st\_large}) supervision, all evaluated on the same dataset with clean test set.\par}
\vspace{-6mm}
\end{table}

\section{MTL Optimization Robustness }
\label{appendix:Optimization_Robustness}

As demonstrated in Table~\ref{tab:optimization-robustness-long} and Table~\ref{tab:optimization-robustness-overall}, the proposed MTL model exhibits strong optimization robustness across both the Long-range and Overall test datasets. While aggressive compiler optimizations (O2 and O3) introduce complex binary transformations that inherently increase the difficulty of the prediction task, the model's performance degradation remains strictly bounded. Across all optimization levels (O0--O3), the model consistently maintains F1 scores above 93\%. Notably, specific instruction types such as \textit{jumptable} show exceptional resilience (maintaining F1 scores near 97\%--98\% even at O3). Although \textit{icall} exhibits a slight sensitivity to higher optimization levels, the overall performance remains highly stable. This consistent accuracy across structural variations confirms the MTL architecture's practical viability for analyzing heavily optimized real-world binaries.
\section{GDH Pruning}
\label{appendix:GDH Pruning}
\begin{table}[h]
\centering
\caption{Ablation study of the Dual Hub architecture with varying GDH neighborhood radii (hops) on long-range ICF pairs.}
\label{tab:dual-hub-long-range-GDH}
\resizebox{0.9\linewidth}{!}{
\begin{tabular}{lccc}
\toprule
\textbf{Setting} & \textbf{Precision (\%)} & \textbf{Recall (\%)} & \textbf{F1 (\%)} \\
\midrule
\multicolumn{4}{c}{\textit{Indirect Call (ic)}} \\
GCH + 1-hop GDH & 88.96 & 91.00 & 89.97 \\
GCH + 2-hop GDH & 90.28 & 92.44 & 91.35 \\
GCH + 3-hop GDH & 90.35 & 92.13 & 91.23 \\
\addlinespace[-0.5mm]
\multicolumn{4}{c}{\textit{Indirect Tail Call (itc)}} \\
GCH + 1-hop GDH & 86.35 & 90.32 & 88.29 \\
GCH + 2-hop GDH & 89.54 & 91.19 & 90.36 \\
GCH + 3-hop GDH & 87.82 & 90.23 & 89.01 \\
\addlinespace[-0.5mm]
\multicolumn{4}{c}{\textit{Jump Table (jt)}} \\
GCH + 1-hop GDH & 92.45 & 94.13 & 93.28 \\
GCH + 2-hop GDH & 95.10 & 95.64 & 95.37 \\
GCH + 3-hop GDH & 93.66 & 95.01 & 94.33 \\
\addlinespace[-0.5mm]
\multicolumn{4}{c}{\textit{Return (ret)}} \\
GCH + 1-hop GDH & 91.51 & 91.03 & 91.27 \\
GCH + 2-hop GDH & 91.60 & 91.16 & 91.38 \\
GCH + 3-hop GDH & 91.04 & 91.28 & 91.16 \\
\bottomrule
\end{tabular}%
}
\vspace{1mm}
\caption*{\textit{Note.} \texttt{GCH + $k$-hop GDH} integrates the Global Data Hub restricted to a $k$-hop radius of code candidates.}
\vspace{-6mm}
\end{table}

\vspace{2pt}
\noindent \textbf{\textit{Impact of GDH neighborhood radius.}} 
To determine the optimal structural context for data nodes, we evaluated the Dual Hub architecture across varying GDH connection radii (Table~\ref{tab:dual-hub-long-range-GDH}). The empirical results reveal a clear trade-off: a restricted 1-hop radius provides insufficient data context, leading to suboptimal predictive performance. Conversely, expanding the radius to 3 hops introduces excessive structural noise from irrelevant data nodes, causing F1 scores to systematically degrade across all task types. The 2-hop radius strikes the optimal balance—it successfully captures essential data references while aggressively pruning irrelevant structural noise, thereby consistently yielding the highest F1 scores. Consequently, we adopt the 2-hop threshold for all GDH configurations.

\section{MTL Performance Without Dual Hubs}
\label{appendix:MTL Performance withoutt Dual Hubs}
\begin{table}[ht]
\centering
\caption{Performance comparison of Single-Task and Multi-Task (MTL) paradigms evaluated \textit{without} Dual Hubs on the \textbf{Long-range test Dataset}. All models were trained using the large-scale static dataset.}
\label{tab:mtl-performance-no-hubs}
\resizebox{0.95\linewidth}{!}{
\begin{tabular}{lccccc}
\toprule
\textbf{Task Type} & \textbf{Setting} & \textbf{Precision (\%)} & \textbf{Recall (\%)} & \textbf{F1 (\%)} & \textbf{\boldmath$\Delta F1$(\%)} \\
\midrule
\multirow{2}{*}{\textit{Indirect Call (ic)}} 
& Single  & 84.73 & 85.89 & 85.31 & -- \\
& MTL     & 87.87 & 86.72 & 87.29 & +1.98 \\
\midrule
\multirow{2}{*}{\textit{Indirect Tail Call (itc)}} 
& Single  & 80.61 & 81.85 & 81.23 & -- \\
& MTL     & 82.75 & 82.87 & 82.81 & +1.58 \\
\midrule
\multirow{2}{*}{\textit{Jump Table (jt)}} 
& Single  & 97.35 & 97.68 & 97.51 & -- \\
& MTL     & 97.05 & 98.01 & 97.53 & +0.02 \\
\midrule
\multirow{2}{*}{\textit{Return (ret)}} 
& Single  & 86.27 & 86.41 & 86.34 & -- \\
& MTL     & 89.16 & 86.99 & 88.06 & +1.72 \\
\bottomrule
\end{tabular}%
}
\vspace{1mm}
\caption*{\textit{Note.} All models in this table are evaluated in the foundational baseline setting (completely without GCH and GDH routing).}
\vspace{-3mm}
\end{table}

As illustrated in Table~\ref{tab:mtl-performance-no-hubs}, in the absence of the Dual Hub architecture, the model achieves a strictly marginal average F1 improvement of only 1.33\%. This pronounced performance plateau empirically exposes a fundamental limitation of standard Graph Neural Networks (GNNs): conventional local message-passing mechanisms suffer from severe multi-hop information decay and over-squashing across extended topological distances. Consequently, without explicit global routing to bridge these structural gaps, standard GNNs
remain bottlenecked by topological unreachability and struggle to capture the long-range dependencies required for complex ICF resolution.



\begin{table}[h]
\centering
\caption{Performance of ICFlowNet on optimization-specific long-range test sets generated by recompiling the same source programs under O0–O3, supervised by a Large Static Dataset. No indirect tail call rows are shown for O0/O1 because these subsets contain no tail-call-eliminated instances.}
\vspace{-2mm}
\label{tab:optimization-robustness-long}
\resizebox{0.43\textwidth}{!}{%
\begin{tabular}{llccc}
\toprule
\thead{\textbf{Op} \\ \textbf{level}} & \textbf{Type} & \textbf{Precision (\%)} & \textbf{Recall (\%)} & \textbf{F1 (\%)} \\
\midrule
\multirow{3}{*}{O0} 
 & icall       & 96.16 & 95.72 & 95.94 \\
 & jumptable   & 98.25 & 97.99 & 98.12 \\
 & ret         & 97.16 & 96.52 & 96.84\\
\midrule
\multirow{3}{*}{O1} 
 & icall       & 94.29 & 95.99 & 95.13 \\
 & jumptable   & 98.28 & 97.70 & 97.99 \\
 & ret         & 95.81 & 96.49 & 96.15\\
\midrule
\multirow{4}{*}{O2} 
 & icall       & 93.78 & 94.36 & 94.07 \\
 & itailcall   & 97.22 & 95.28 & 96.24 \\
 & jumptable   & 98.11 & 95.99 & 97.04 \\
 & ret         & 95.39 & 96.17 & 95.78 \\
\midrule
\multirow{4}{*}{O3} 
 & icall       & 93.46 & 93.88 & 93.67 \\
 & itailcall   & 96.82 & 95.43 & 96.12 \\
 & jumptable   & 98.11 & 95.68 & 96.88 \\
 & ret         & 95.08 & 94.98 & 95.03 \\
\bottomrule
\end{tabular}%
}
\vspace{-3mm}
\end{table}

\begin{table}[h]
\centering
\caption{Performance of ICFlowNet on optimization-specific overall test sets generated by recompiling the same source programs under O0–O3, supervised by a Large Static Dataset. No indirect tail call rows are shown for O0/O1 because these subsets contain no tail-call-eliminated instances.}
\vspace{-2mm}
\label{tab:optimization-robustness-overall}
\resizebox{0.43\textwidth}{!}{%
\begin{tabular}{llccc}
\toprule
\thead{\textbf{Op} \\ \textbf{level}} & \textbf{Type} & \textbf{Precision (\%)} & \textbf{Recall (\%)} & \textbf{F1 (\%)} \\
\midrule
\multirow{3}{*}{O0} 
 & icall       & 98.49 & 96.00 & 97.23 \\
 & jumptable   & 99.11 & 98.13 & 98.62 \\
 & ret         & 98.28 & 96.42 & 97.34\\
\midrule
\multirow{3}{*}{O1} 
 & icall       & 97.79 & 97.13 & 97.46 \\
 & jumptable   & 98.31 & 97.93 & 98.12 \\
 & ret         & 97.85 & 96.64 & 97.24\\
\midrule
\multirow{4}{*}{O2} 
 & icall       & 95.72 & 94.98 & 95.35 \\
 & itailcall   & 97.53 & 96.44 & 96.98 \\
 & jumptable   & 98.42 & 97.64 & 98.03 \\
 & ret         & 97.32 & 95.89 & 96.60 \\
\midrule
\multirow{4}{*}{O3} 
 & icall       & 94.52 & 94.06 & 94.29 \\
 & itailcall   & 96.85 & 95.99 & 96.42 \\
 & jumptable   & 98.05 & 97.63 & 97.84 \\
 & ret         & 96.23 & 95.35 & 95.79 \\
\bottomrule
\end{tabular}%
}
\vspace{-2mm}
\end{table}

\section{Feature Level Fusion vs. MTL}
\label{appendix:Feature level fusion}

\vspace{2pt}
\noindent
\textbf{\textit{Why simple feature fusion is insufficient.}}
To assess whether the gains of our multi-task design could be reproduced by ensemble-style alternatives, we first note that simple ensemble strategies (e.g., majority voting or probability averaging) only aggregate the predictions of independently trained single-task models and therefore do not provide cross-task structural awareness.
A similar limitation also applies to simple feature-fusion schemes, which combine task-specific representations in a shallow manner but still lack joint task-level optimization. As shown in Table~\ref{tab:mtl-performance-fusion}, \textit{Feat-fusion} yields only limited and inconsistent improvements over the Single-Task baseline. In particular, it provides a relatively larger gain on the \textit{return} task ($+4.61$ F1), which suggests that simple fusion can be more effective for task types with a larger number of supervision pairs, where richer training signals may help stabilize the fused representation. However, this advantage does not generalize to lower-resource task types. For tasks with fewer supervision pairs, the fusion process may introduce undesirable side effects, such as feature interference, noise propagation, or negative transfer across tasks. This is reflected by the performance drops on \textit{itc} ($-1.57$) and \textit{jt} ($-0.62$), and by the much smaller gain on \textit{ic} ($+1.37$). In contrast, the MTL variant with auxiliary type prediction (\textit{MTL\_t}) consistently improves over the single-task baseline on all task types, with gains of $+4.48$, $+5.13$, $+0.44$, and $+4.91$ F1 on \textit{ic}, \textit{itc}, \textit{jt}, and \textit{ret}, respectively. On average, \textit{Feat-fusion} improves F1 by only $0.95$ points over the Single-Task baseline, whereas \textit{MTL\_t} achieves an average gain of $3.74$ points. These results indicate that na\"{\i}vely combining features from isolated models may offer some benefit for data-rich tasks, but it is not a reliable solution overall. By contrast, jointly optimized MTL can better exploit shared structural regularities while avoiding the side effects of shallow fusion, resulting in more robust improvements across all task types.

\begin{table}[h]
\centering
\caption{Performance comparison of Single-Task, Feat-fusion, and the auxiliary type-prediction variant (MTL\_t) on the \textit{Long-range test dataset}. All models are trained on the large-scale static dataset. \textit{Note:} \textsuperscript{\dag}Dual hub setting; \textsuperscript{\ddag}No hub setting.}
\vspace{-2mm}
\label{tab:mtl-performance-fusion}
\resizebox{0.95\linewidth}{!}{
\begin{tabular}{lccccc}
\toprule
\textbf{Task Type} & \textbf{Setting} & \textbf{Precision (\%)} & \textbf{Recall (\%)} & \textbf{F1 (\%)} & \textbf{\boldmath$\Delta F1$(\%)} \\
\midrule
\multirow{3}{*}{\textit{Indirect Call (ic)}} 
& Single\textsuperscript{\dag}  & 90.28 & 92.44 & 91.35 & -- \\
& Feat-fusion                  & 92.51 & 92.93 & 92.72 & +1.37 \\
& MTL\_t                       & \textbf{95.36} & 96.30 & \textbf{95.83} & \textbf{+4.48} \\
\midrule
\multirow{3}{*}{\textit{Indirect Tail Call (itc)}} 
& Single\textsuperscript{\dag}  & 89.54 & 91.19 & 90.36 & -- \\
& Feat-fusion                  & 88.25 & 89.34 & 88.79 & -1.57 \\
& MTL\_t                       & 95.81 & 95.17 & 95.49 & +5.13 \\
\midrule
\multirow{3}{*}{\textit{Jump Table (jt)}} 
& Single\textsuperscript{\ddag} & 97.35 & 97.68 & 97.51 & -- \\
& Feat-fusion                  & 96.36 & 97.42 & 96.89 & -0.62 \\
& MTL\_t                       & \textbf{98.25} & 97.65 & \textbf{97.95} & \textbf{+0.44} \\
\midrule
\multirow{3}{*}{\textit{Return (ret)}} 
& Single\textsuperscript{\dag}  & 91.60 & 91.16 & 91.38 & -- \\
& Feat-fusion                  & 95.78 & 96.21 & 95.99 & +4.61 \\
& MTL\_t                       & 96.11 & 96.47 & 96.29 & +4.91 \\
\bottomrule
\end{tabular}%
}
\vspace{-3mm}
\end{table}

\begin{table}[t]
\centering
\caption{External comparison under BinDSA's profiling-based precision/recall protocol~\cite{Gao2025}.}
\vspace{-2mm}
\label{tab:bindsa_comp}
\resizebox{0.9\columnwidth}{!}{%
\begin{tabular}{@{}lcccc@{}}
\toprule
\multirow{2}{*}{\textbf{Program}} & \multicolumn{2}{c}{\textbf{BinDSA}} & \multicolumn{2}{c}{\textbf{ICFlowNet}} \\ \cmidrule(lr){2-3} \cmidrule(l){4-5}
 & \textbf{Precision} & \textbf{Recall} & \textbf{Precision} & \textbf{Recall} \\ \midrule
401.bzip2     & 1.00 & 1.00 & 1.00 & 1.00 \\
458.sjeng     & 0.86 & 1.00 & 0.92 & 1.00 \\
433.milc      & 1.00 & 1.00 & 1.00 & 1.00 \\
456.hmmer     & 0.91 & 1.00 & 0.92 & 1.00 \\
464.h264ref   & 0.65 & 1.00 & 0.95 & 1.00 \\
445.gobmk     & 0.56 & 0.95 & 0.92 & 1.00 \\
400.perlbench & 0.53 & 0.86 & 0.94 & 0.99 \\
403.gcc       & 0.46 & 0.67 & 0.90 & 0.97 \\ \bottomrule
\end{tabular}%
}
\end{table}

\begin{table}[t]
\centering
\caption{External comparison under the BPADisa AICT/recall convention of Disa~\cite{Wang2025}.}
\vspace{-2mm}
\label{tab:disa_comp}
\resizebox{0.97\columnwidth}{!}{%
\begin{tabular}{@{}lrcccc@{}}
\toprule
\multirow{2}{*}{\textbf{Program-Opt}} & \multirow{2}{*}{\textbf{\# Csite}} & \multicolumn{2}{c}{\textbf{BPADisa}} & \multicolumn{2}{c}{\textbf{ICFlowNet}} \\ \cmidrule(lr){3-4} \cmidrule(l){5-6}
 &  & \textbf{AICT} & \textbf{Recall} & \textbf{AICT} & \textbf{Recall} \\ \midrule
hmmer-00     & 9   & 2.8   & 100  & 2.4   & 100  \\
hmmer-01     & 11  & 4.3   & 100  & 3.9   & 100  \\
hmmer-02     & 10  & 2.8   & 100  & 2.5   & 100  \\
hmmer-03     & 9  & 1.0   & 100  & 2.5   & 100  \\ \midrule
h264ref-00   & 369 & 4.3   & 100  & 3.6   & 100  \\
h264ref-01   & 353 & 5.2   & 99.7 & 3.9   & 100  \\
h264ref-02   & 352 & 26.9  & 100  & 18.3  & 100  \\
h264ref-03   & 355 & 18.6  & 100  & 18.2  & 99.5 \\ \midrule
gobmk-00     & 44  & 846.9 & 100  & 532.3 & 100  \\
gobmk-01     & 44  & 1336.3 & 100 & 1218.2 & 100 \\
gobmk-02     & 44  & 1337.7 & 100 & 1292.1 & 100 \\
gobmk-03     & 44  & 1416.2 & 100 & 1272.7 & 100 \\ \midrule
perlbench-00 & 139 & 387.6 & 99.1 & 328.6 & 99.6 \\
perlbench-01 & 139 & 379.7 & 100  & 317.3 & 99.6 \\
perlbench-02 & 110 & 373.9 & 100  & 313.2 & 98.4 \\
perlbench-03 & 237 & 456.4 & 100  & 364.3 & 98.2 \\ \midrule
gcc-00       & 459 & 447.1 & 95.3 & 327.2 & 95.3 \\
gcc-01       & 473 & 338.1 & 95.8 & 351.4 & 95.7 \\
gcc-02       & 450 & 431.1 & 96.2 & 381.3 & 97.1 \\
gcc-03       & 727 & 454.1 & 96.2 & 392.9 & 96.7 \\ \bottomrule
\end{tabular}%
}
\end{table}

\section{External Comparison}
\label{appendix:External_Comparison}

To complement the main-text comparison against learned baselines, we also report \emph{external} comparisons to recent program-analysis systems under their published benchmark conventions. These results should not be interpreted as strict artifact-level reproductions: BinDSA~\cite{Gao2025} reports profiling-based precision/recall on a SPEC-style benchmark, while Disa~\cite{Wang2025} evaluates indirect-call refinement through the BPADisa pipeline using AICT and recall across optimization levels. We therefore compare against the same benchmark families and metrics used in those papers, and use the results as cross-framework reference points rather than exact apples-to-apples reproductions.

\vspace{2pt}
\noindent
\textbf{\textit{Comparison to BinDSA.}}
Table~\ref{tab:bindsa_comp} compares ICFlowNet with BinDSA~\cite{Gao2025} under BinDSA's profiling-based precision/recall convention. On simpler programs such as \textit{401.bzip2} and \textit{433.milc}, both systems are essentially perfect. On larger and more structurally complex binaries, however, ICFlowNet maintains substantially higher precision while keeping recall comparable or better. The difference is most pronounced on \textit{403.gcc}, where BinDSA reports precision/recall of 0.46/0.67, while ICFlowNet reaches 0.90/0.97. Similar trends appear on \textit{445.gobmk} and \textit{400.perlbench}. These results suggest that learned structural representations remain robust even when precise formal reasoning becomes harder to scale.

\vspace{2pt}
\noindent
\textbf{\textit{Comparison to BPADisa.}}
Table~\ref{tab:disa_comp} compares ICFlowNet with BPADisa, i.e., the Disa-facilitated BPA~\cite{BPA} pipeline evaluated in~\cite{Wang2025}. Here the comparison metric is AICT (lower is better) together with recall. Across optimization levels, ICFlowNet generally yields smaller indirect-call candidate sets while preserving similar recall. The reductions are especially visible on larger optimized binaries: for example, ICFlowNet reduces AICT from 456.4 to 364.3 on \textit{perlbench-03}, from 454.1 to 392.9 on \textit{gcc-03}, and from 1416.2 to 1272.7 on \textit{gobmk-03}, while maintaining nearly identical recall. This indicates that ICFlowNet can tighten target sets without relying on iterative block-memory analysis.

\vspace*{2pt}
\noindent
\textbf{\textit{Inference-time cost profile.}}
These systems differ not only in accuracy, but also in deployment cost. BinDSA~\cite{Gao2025} performs context-sensitive heap reconstruction and pointer reasoning, while BPADisa~\cite{Wang2025} still inherits BPA-style block-memory analysis. Their runtime therefore grows from minutes to hours on large binaries. For example, on \texttt{gcc}, BPA reports $27,619$ seconds and BinDSA reports $1,865$ seconds~\cite{BPA,Gao2025}. In contrast, ICFlowNet amortizes model fitting into an offline training phase. At deployment time, inference reduces to a single feed-forward pass over the recovered binary graph, taking about 25 seconds of end-to-end wall-clock time on the same \texttt{gcc} binary. This lighter inference-time cost makes ICFlowNet substantially more practical for whole-system downstream analyses where formal program-analysis pipelines may time out. 

\end{document}